\documentclass[11pt,a4paper, pdf]{article}

\def\lamqcd{{$\Lambda_{\mrm{QCD}}$}}

\usepackage{graphicx}
\usepackage{hyperref}
\usepackage{xspace}
\usepackage{cite}

\newcommand{\mrm}[1]{\ensuremath{\mathrm{#1}}}
\newcommand{\ttt}[1]{\texttt{#1}}

\usepackage{amsmath}

        \newdimen\eqskip
        \newdimen\txtskip
        \baselineskip=\txtskip
        \newdimen\mysep                
        \newdimen\hmysep
\def\be{\begin{equation}}
\def\ee{\end{equation}}

\newcommand\sss{\scriptscriptstyle\rm}
\newcommand\pt{p_{\sss T}\xspace}
\newcommand\ptsq{p_{\sss T}^2\xspace}

\def    \=              {\;=\;} 

\usepackage{cite}
\usepackage{graphicx}
\usepackage{mathptmx}
\usepackage{xspace}
\usepackage{xcolor}

\usepackage{longtable}
\usepackage{threeparttable}
\usepackage{booktabs}
\usepackage{multicol}
\usepackage{multirow}
\usepackage[small,nooneline,center]{subfigure}
\usepackage[italic]{hepparticles}
\usepackage{hepnames,hepunits}

\usepackage{setspace}
\usepackage{threeparttable}
\definecolor{todoboxcolor}{rgb}{1.0,0.6,0.6}

\definecolor{todoneboxcolor}{rgb}{1.0,0.9,0.9}

\newlength{\capindent}
\newlength{\capwidth}
\newlength{\figwidth}
\newcommand{\icaption}[2][!*!,!]{\hspace*{\capindent}%
  \begin{minipage}{\capwidth}
    \ifthenelse{\equal{#1}{!*!,!}}%
      {\caption{#2}}%
      {\caption[#1]{#2}}
      \vspace*{3mm}
  \end{minipage}}
\def\pt{p_{\sss T}\xspace}
\def\ptsq{p_{\sss T}^2\xspace}
\def\jimmy{{Jimmy}\xspace}
\def\herwig{{Herwig}\xspace}

\def\pythia{{Pythia}\xspace}
\def\madgraph{{MadGraph}\xspace}
\def\professor{{Professor}\xspace}
\def\rivet{Rivet\xspace}

\def\ariadne{Ariadne\xspace}

\def\rivet{Rivet\xspace}
\def\professor{Professor\xspace}

\def\professor{Professor\xspace}
\def\rivet{Rivet\xspace}

\def\pythiasix{Pythia~6\xspace}

\def\herwig{Herwig\xspace}

\def\alpgen{AlpGen\xspace}

\def\alpgenpythiasix{AlpGen~+~Pythia~6\xspace}

\def\pythiasixstand{Pythia~6 standalone\xspace}

\def\jimmy{Jimmy\xspace}
\def\rivet{Rivet\xspace}

\def\professor{Professor\xspace}

\def\alphaS{\ensuremath{\alpha_S}\xspace}
\def\LambdaQCD{\ensuremath{\Lambda_\text{QCD}}\xspace}

\def\pzero{Perugia~0\xspace}
\def\psoft{Perugia~Soft\xspace}
\def\phard{Perugia~Hard\xspace}
\def\pzehn{Perugia~2010\xspace}
\def\pelf{Perugia~2011\xspace}
\def\pelfradhi{Perugia~2011~radHi\xspace}
\def\pelfradlo{Perugia~2011~radLo\xspace}
\def\pelfmatched{Perugia~2011~``matched''\xspace}
\def\xlclu{\ttt{xlclu}\xspace}

\def\ktfac{\ttt{ktfac}\xspace}
\begin{document}
\begin{titlepage}
{\begin{flushright}{
 \begin{minipage}{5cm}
   CERN-PH-TH/2011-228\\
   DESY 11-124 \\
\end{minipage}}\end{flushright}}
\vspace*{2cm}
\begin{center}
{\Large\textbf{Monte Carlo tuning in the presence of Matching}}\\[0.9cm]
B.~Cooper$^1$, J.~Katzy$^2$, M.~L.~Mangano$^3$, A.~Messina$^3$,
L.~Mijovi\'c$^2$, P.~Skands$^3$\\[5pt] 
\end{center}\vspace*{5mm}
{\small $^1$ Department of Physics \& Astronomy, University College London, WC1E 6BT, UK\\[5pt]
$^2$ Deutsches Electron Synchroton, 22603 Hamburg, Germany \\[5pt] 
$^3$ European Organization for Nuclear Research, CERN CH-1211,
Geneva 23, Switzerland\\[5pt]}
\vspace*{0.3cm}
\vspace*{0.5cm}
\begin{abstract}
 We consider the impact of varying $\alpha_s$ choices (and scales) on
each side of the so-called ``matching scale'' in MLM-matched matrix-element +
parton-shower predictions of collider  
observables. We explain how inconsistent prescriptions can lead to
counter-intuitive results and present a few explicit examples, 
focusing mostly on $W/Z$ + jets processes. We give a specific 
prescription for how to improve the consistency of the matching and also
address how to perform consistent tune variations (e.g., of the
renormalization scale) around a central choice. Comparisons to several
collider processes are included to illustrate the properties of the
resulting improved matching, relying on \alpgenpythiasix, with the
latter using the so-called Perugia 2011 tunes, developed as part of
this effort. 
\end{abstract} 

\end{titlepage}


\section{Introduction}
The theoretical description of multijet production in hadronic
collisions is one of the key ingredients for the interpretation of the
data from high-energy hadron colliders, the 1.96 TeV proton-antiproton
Tevatron collider at Fermilab, and the proton-proton Large Hadron
Collider (LHC) at CERN. Final states with multijets, possibly
associated with electroweak gauge bosons, are in fact the dominant
signature of the decay of heavy particles produced at high energy,
whether in the Standard Model (top quarks and Higgs bosons), or in
theories beyond the Standard Model (BSM), such as supersymmetry. The
identification of these particles, and the study of their properties,
requires an accurate modelling of the Standard Model (SM) sources of
multijets. Great progress was achieved towards this goal in the past
decade. On one side, the calculation of inclusive, parton-level,
cross-sections to next-to-leading-order (NLO) in QCD has produced
results for processes as complex as W+4 jets \cite{Berger:2010zx}. 
On the other, algorithms
have been developed and implemented in numerical codes to provide a
 complete description of the hadronic final states emerging from
 processes with up to 6 jets, merging the exact leading-order (LO)
calculation of the partonic matrix elements (ME) with the evolution,
provided by so-called shower Monte Carlo (MC) codes, of the partonic
shower (PS) and the subsequent hadronization of the partons in to physical
hadrons. 

The development of theoretical tools has been accompanied by
experimental measurements, which provide the necessary validation
test-bed for these calculations. Parton-level NLO calculations provide
a first-principle description of inclusive final states: they have an
intrinsic high degree of precision, due to the reduced dependence on
the unphysical choice of a renormalization and factorization scale
and, furthermore, are not subject to modelling uncertainties related
to the details of the non-perturbative phase of the final state
evolution. Calculations based on the merging of LO matrix elements,
shower evolution and hadronization, on the other hand, while affected
by the larger scale-setting uncertainty due to the LO approximation,
provide a fully exclusive description of the final states, and are
therefore more suitable for the experimental analyses. Their ultimate
goal is not only to give reliable estimates of the inclusive jet rates
and energy distributions, but also to reproduce properties of the
final states such as the jet inner structure and the distribution of
softer particles produced outside of the jets, including those
resulting from the evolution of the fragments of the original
colliding hadrons\footnote{We refer to the ensemble of these particles
as the “underlying event”, or UE}. These properties, which depend on
the details of the non-perturbative dynamics, can only be described
through the phenomenological models embedded in the shower MC
codes. The parameters of these phenomenological models need to be tuned
using experimental data of some suitable observables. The
factorization assumption built into any description of large-Q
processes justifies the use of these same parameters in the prediction
of different observables, and provides the basis for the predictive
power of such tools. This assumption however must be validated with a
direct comparison with data. Elements that need to be probed include
the scaling with beam energy of the UE parameters, the universality of
the parameters controlling the shower evolution and hadronization, and
the overall independence of all parameters on the type of hard
process. Deviations from the expected universality would highlight
faults in the underlying modelling of effects beyond perturbative
physics, or could be due to the insufficient precision of the
perturbative description, in case NLO effects were to modify
significantly the LO predictions. Differences compatible with the
theoretical systematics of the LO approximation could however be
reabsorbed by modifying the perturbative parameters that govern the LO
systematics, for example the renormalization and factorization scales,
or the matching variables used in the matrix-element/shower merging
algorithm. 

It is therefore important to understand the correlations
between the effects of changing the soft and UE parameters on one
side, and the perturbative parameters on the other.  In this paper we
present studies which demonstrate that, in the tuning of ME-PS matched
predictions, it is vital that there is consistency in the treatment of
\alphaS~in both the ME and PS components. While this is a general
issue for all shower MCs, we consider as an explicit example
the merging of LO matrix elements
with the \pythiasix shower MC~\cite{Sjostrand:2006za}, 
as implemented in the framework of the
\alpgen code~\cite{Mangano:2002ea}, one of the reference tools for
experimental multijet 
studies at the Tevatron and at the LHC. The most recent versions of
\pythia~(6.425) and \alpgen~(2.14) codes were used for  
producing the results. 

On the \pythiasix side, we consider several different tune variations of
the interleaved $\pt$-ordered parton-shower model \cite{Sjostrand:2004ef},
focusing on the so-called ``Perugia'' set of tunes of
\cite{Skands:2009zm,Skands:2010ak}, ranging from the Perugia 0
tune (from 2009) to the Perugia 2011 updates that have been developed as part of
this work, including  systematic up/down variations of the shower
activity (see the Appendix and \cite{Skands:2009zm,Skands:2010ak} for
details). We also compare to the ``DW'' tune~\cite{Albrow:2006rt} of the virtuality-ordered shower model
\cite{Sjostrand:1985xi,Bengtsson:1986et}. For
\herwig~\cite{Corcella:2000bw}, 
we include the ``Jimmy'' underlying-event model \cite{Jimmy}, with default
parameters. We emphasize that 
the qualitative conclusions presented in this paper carry over to
other shower models, including the ones implemented in 
\pythia~8~\cite{Corke:2010zj,Corke:2010yf} and \herwig++~\cite{Bahr:2008pv}, 
but the quantitative
aspects should still be considered limited to the particular tunes 
and shower models studied here. We rely on Fastjet \cite{Cacciari:2005hq} 
for jet clustering and have further used the \rivet-based~\cite{Buckley:2010ar} 
mcplots web site \cite{mcplots} for some of our comparisons.

The paper is organised as follows: in Section~\ref{sec:consistent} we describe in detail the theoretical
nature of the \alphaS~consistency problem, and give a practical example of how it can
be manifest in the prediction of high $\pt$~observables. In
Section~\ref{sec:stabalising} we show how a simple prescription can be
applied to stabilise ME-PS tunings against this problem, propose
a new tune for \alpgenpythiasix matched predictions, and demonstrate
the behaviour of this tune under tuning variations. In
Section~\ref{sec:datacomp} we show that this new \alpgenpythiasix tune
is able to reproduce (within statistical errors) the Tevatron and LHC
vector boson plus jets data. In addition we also the tune predictions to the jet shape measurements at
the Tevatron and LHC. Finally we conclude in Section~\ref{sec:conclusion}.

\section{The Importance of Consistent \alphaS Treatment in ME-PS Matched Predictions\label{sec:consistent}}

In this section we demonstrate that consistent treatment of \alphaS~in
ME-PS matched predictions is important in order to achieve the desired
accuracy in the prediction of high $\pt$~observables. We first present the theoretical
arguments behind this, and then go on to show and explain that without adopting
this approach one can observe undesirable and counter-intuitive effects on
experimental observables. 

\subsection{Theoretical Background\label{sec:theory}}
The philosophy behind matching prescriptions such as the MLM one
\cite{Mangano:2006rw,Mrenna:2003if} 
employed by \alpgen is to separate phase space cleanly into two distinct
regions; a short-distance one, which is supposed to be described by matrix elements,
and a long-distance one described by parton showers.  
In the long-distance region,  
real and virtual corrections, with the latter represented by Sudakov
factors, are both generated by the shower and are intimately related
by unitarity (for pedagogical reviews, see, e.g., \cite{Buckley:2011ms,Skands:2011pf}). 
On the short-distance side, the real corrections are
generated by the matrix elements while the virtual ones are still
generated by the shower. 

Much effort has gone into ensuring that the behaviour across the
boundary between the two regions be as
smooth as possible. CKKW showed \cite{Catani:2001cc} that it is
possible to remove any  dependence on this ``matching scale'' at NLL
precision by careful choices of all ingredients in the matching;
technical details of the implementation 
are important, and the dependence on the unphysical matching scale
may be larger than NLL unless the implementation
matches the theoretical algorithm
precisely~\cite{Lonnblad:2001iq,Lavesson:2005xu,Lavesson:2008ah}. 

Especially when two different computer codes are used for matrix
elements and showering, respectively (as when \alpgen 
or
\madgraph~\cite{Alwall:2007st} 
is combined with \pythiasix 
or \herwig 
),
inconsistent parameter 
sets between the two codes can jeopardise the consistency of the
calculation and lead to unexpected results, as will be illustrated in
the following sections.

To give a very simple theoretical example, suppose a matched
matrix-element generator (MG) uses a different definition of
$\alpha_s$ than the parton-shower generator (SG). Suppressing parton
luminosity factors to avoid clutter, the real corrections, integrated
over the hard part of phase space, for some arbitrary final state $F$,
will then have the form
\begin{equation}
\sigma_{F+1}^{\mrm{incl}} = \int_{Q_F^2}^s \mrm{d}\Phi_{F+1}
\ \alpha_s^{\mrm{MG}} \ |M_{F+1}|^2~, 
\end{equation}
where we have factored out the coupling corresponding to the ``+1''
parton and suppressed the dependence on 
any other couplings that may be present in $|M_{F+1}|^2$.
The virtual corrections at the same order, 
generated by the shower off $F$, will have the form
\begin{equation}
\sigma_{F}^{\mrm{excl}} = \sigma_{F}^{\mrm{incl}} - \int
\mrm{d}\Phi_{F} \int_{Q_F^2}^{s} \frac{\mrm{d}Q^2}{Q^2} \  
\mrm{d}z \ \sum_i \frac{\alpha_s^{\mrm{SG}}}{2\pi}P_i(z) \ |M_{F}|^2
\ + \
{\cal O}(\alpha_s^2)~, \label{eq:virtual}
\end{equation}
with  $P_i(z)$ the 
DGLAP splitting kernels (or equivalent radiation
functions in dipole or antenna shower approaches). 
If the two codes use the same definitions for the strong coupling,
$\alpha_s^{\mrm{SG}}=\alpha_s^{\mrm{MG}}$, then the fact that
$P(z)/Q^2$ captures the leading singularities of $|M_{F+1}|^2$
guarantees that the difference between the two expressions can at most
be a non-singular term. Integrated over phase space, such a term 
merely leads to a finite ${\cal O}(\alpha_s)$ change to the total
cross section, which is within the expected precision. 
Indeed, it is a central ingredient in both the MLM
and (L)-CKKW matching
prescriptions that a reweighting of the
matched matrix elements be performed in order to ensure that 
the scales appearing in $\alpha_s$ match
smoothly between the hard and soft regions. Thus, we
may assume that the choice of renormalization scale after matching 
is $\mu \sim \pt$ on both sides of the matching
scale, where $\pt$ is a scale characterising the momentum transfer
at each emission vertex, as established by \cite{Amati:1980ch,Catani:1990rr} 
and encoded in the CKKW formalism~\cite{Catani:2001cc}. 

In the case of the CKKW approach as implemented in the Sherpa MC
framework~\cite{Gleisberg:2008ta}, this 
prescription can be controlled exactly, since the 
matrix element and the shower evolution are part of the same computer
code and hence naturally use the same $\alpha_s$ definition.
This is also true in L{\"o}nnblad's variant
  \cite{Lonnblad:2001iq} of the algorithm, used in 
  \ariadne~\cite{Lonnblad:1992tz}. 
In the case of codes like AlpGen or Madgraph, on the
other hand, an issue emerges. These codes are designed to generate
parton-level event samples to be used with an arbitrary shower
MC. Different shower MCs however use slightly different scales for the
parton branchings, as a result of different approaches to the shower
evolution, and may use different values of \LambdaQCD, as a
result of the tuning of the showers and/or underlying events. 
A possible mismatch
therefore arises in the values of $\alpha_s$ used by the
matrix-element calculation and those used by the shower.

If there is a mismatch in  \LambdaQCD or 
$\alpha_s(M_Z)$, then this will effectively generate a real-virtual
difference whose leading singularities are proportional to 
\begin{equation}
\alpha_s^2 \ b_0\ln\left(
  \frac{\Lambda^2_{\mrm{MG}}}{\Lambda^2_{\mrm{SG}}}
\right) \frac{\mrm{d}Q^2}{Q^2} \ \sum_i P_i(z) \ |M_{F}|^2~.
\end{equation}
which is of next-to-leading logarithmic order (unless
$\Lambda_{\mrm{MG}}\sim\Lambda_{\mrm{SG}}$, in which case it vanishes).
Similarly, even if both matrix-element and shower 
codes are using the same \LambdaQCD, but
they use different running orders, then there will be an
${\mathcal{O}}(\alpha_s^3 \ln(\ptsq/\Lambda^2))$ mismatch, 
which may also become large if $\pt \gg \Lambda$.

To be more concrete, let us consider a specific example. Compare A) a
matched MG+SG calculation which uses the same \LambdaQCD
value on both sides of the matching to B) a calculation in 
which the value used on the MG side is reduced to half its previous
value but the SG one remains the same, as summarised by the two first
columns of tab.~\ref{tab:lambdas}.  
\begin{table}[t]
\begin{center}
\begin{tabular}{lccc}
\toprule
 & A & B & C \\
\midrule
$\Lambda_{\mrm{MG}}$ & $\Lambda$ & $\frac12\Lambda$ & $\Lambda$ \\
$\Lambda_{\mrm{SG}}$ & $\Lambda$ & $\Lambda$ & $\frac12\Lambda$\\
\bottomrule
\end{tabular}
\caption{The three cases, A, B, and C discussed
  in the text, for an arbitrary reference $\Lambda$ value.
\label{tab:lambdas}
}
\end{center}
\end{table}
Going from case A to B, the
following changes result:
\begin{enumerate}
\item The number of $(F+1)$ states added by the MG decreases, due to
  the lowering of the  \LambdaQCD value on the MG side, while
  the number of surviving $F$ states remains constant, since the shower
  Sudakov is not modified. The total estimated cross section therefore
  decreases. 
\item At the differential level, the smaller number of $(F+1)$
  states combined with the unchanging number of $F$ states 
  implies smaller absolute jet cross sections and smaller fractions 
  $\sigma_{\mrm{jet}}/\sigma_{\mrm{tot}}$.
\end{enumerate}
Similarly we may consider what happens if C) we reduce the
\LambdaQCD 
value on the SG side instead, as summarised in the last column of
tab.~\ref{tab:lambdas}. Going from case A to C, the following changes
result:
\begin{enumerate}
\item The number of $(F+1)$ states added by the MG remains constant,
  while the number of surviving $F$ states increases, since the SG is
  generating fewer branchings. The total estimated cross section
  therefore \emph{increases}.
\item Since the number of $(F+1)$ states is constant, while the shower
  is made less active, the final jets will actually be narrower, which
  \emph{increases} the rate of reconstructed jets at any given fixed 
  $\pt$ value. 
\item Since both the total cross section increases and the number of
  reconstructed additional jets also increases, jet \emph{fractions}
  can either increase or decrease.
\end{enumerate}
In particular, note the somewhat counter-intuitive effect that
\emph{decreasing} the shower $\alpha_s$ value actually
\emph{increases} the jet rates in a matched calculation, while it
normally \emph{decreases} them in a standalone shower calculation. 

Since, as was discussed above, inconsistencies among the choices on
the two sides can lead to differences at the NLL level, it is
obviously important to ensure that they are consistent within a
reasonable margin. This is particularly true in the context of
event-generator tuning, in which specifically the NLL components of
the shower description are sought to be optimized with respect to
measured data, and hence changes at this level could effectively
destroy the tuning. 

Finally, we remind the reader that a change in
\LambdaQCD can be interpreted as a change in the 
opposite direction of the renormalization scale argument 
(for constant \LambdaQCD), modulo
small flavour threshold effects that we shall ignore here. This is
easy to realise from the definition of the coupling, 
\begin{equation}
\alpha_s(k \mu^2) \ \stackrel{\mrm{1-loop}}{=} \ \frac{1}{b_0\ln\left(k\mu^2/\Lambda^2\right)}  
\ = \ \frac{1}{b_0\ln\left(\mu^2/(\frac{1}{k}\Lambda^2)\right)}~.
\label{eq:alphas}
\end{equation}
Thus, we may write renormalization scale variations (e.g., by a factor of 2 in
each direction) either by applying a prefactor directly on the renormalization
scale argument of $\alpha_s$ or by applying the inverse of that factor
to \LambdaQCD while keeping the renormalization scale
argument unchanged. Due to the technical structure of the 
codes, the former is more convenient in  \alpgen (via the \ktfac
setting) whilst the latter is
more convenient for \pythiasix.

\subsection{Examples of the interplay between tunes and matching\label{sec:example}}

In this section we give several examples of how the issues in ME-PS matching 
described in Section~\ref{sec:consistent} can affect high-$\pt$~observables using
\alpgen~interfaced to \pythiasix with DW~\cite{Albrow:2006rt}, \pzero
(P0) \cite{Skands:2009zm,Bartalini:2010su} and \pzehn (P2010) tunes
\cite{Skands:2010ak}. 

\begin{figure}
\begin{center}
\begin{minipage}{150mm}
\includegraphics[width=0.85\textwidth]{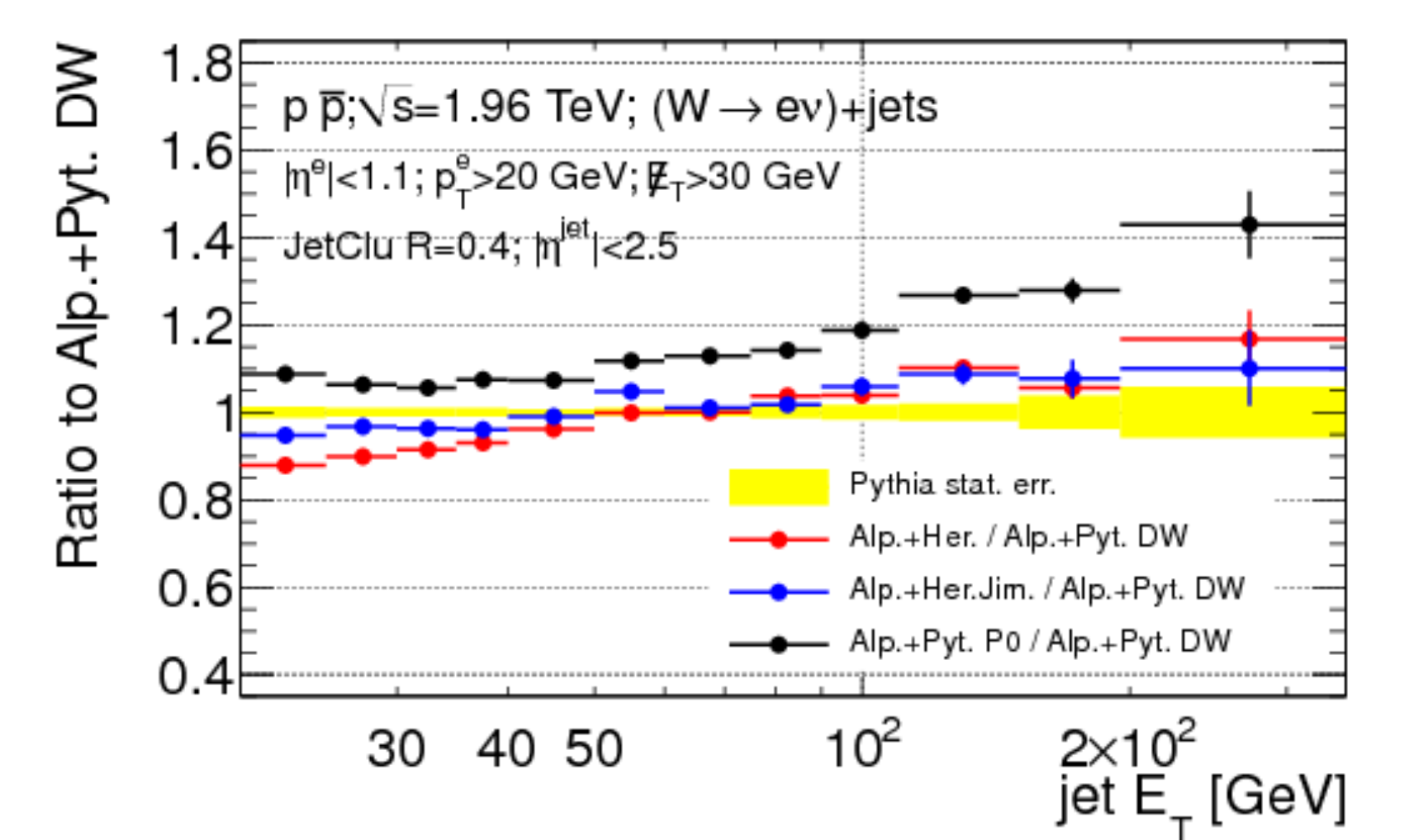}
\caption{Ratio of predictions for the leading-jet $E_T$ spectrum in W+jets final states at
  the Tevatron, obtained with \alpgen plus various MC codes and
  tunes. The leading jet observable is defined at the particle-level
  as in the CDF W+jets analysis~\cite{Aaltonen:2007ip}.}
\label{fig:wjets_MCcomp}
\end{minipage}
\end{center}
\end{figure}

\begin{figure}
\begin{center}
\begin{minipage}{150mm}
\includegraphics[width=0.5\textwidth]{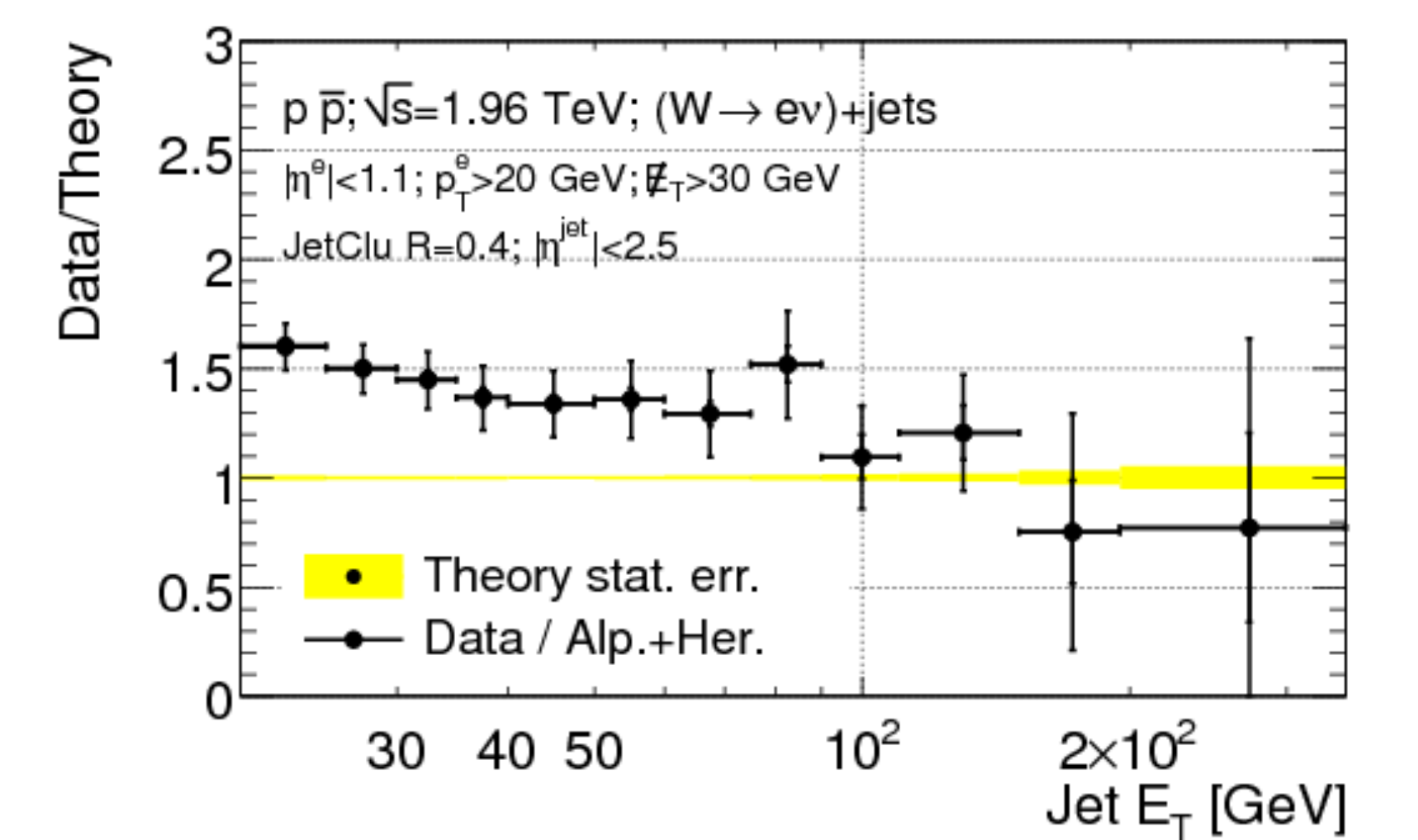}
\includegraphics[width=0.5\textwidth]{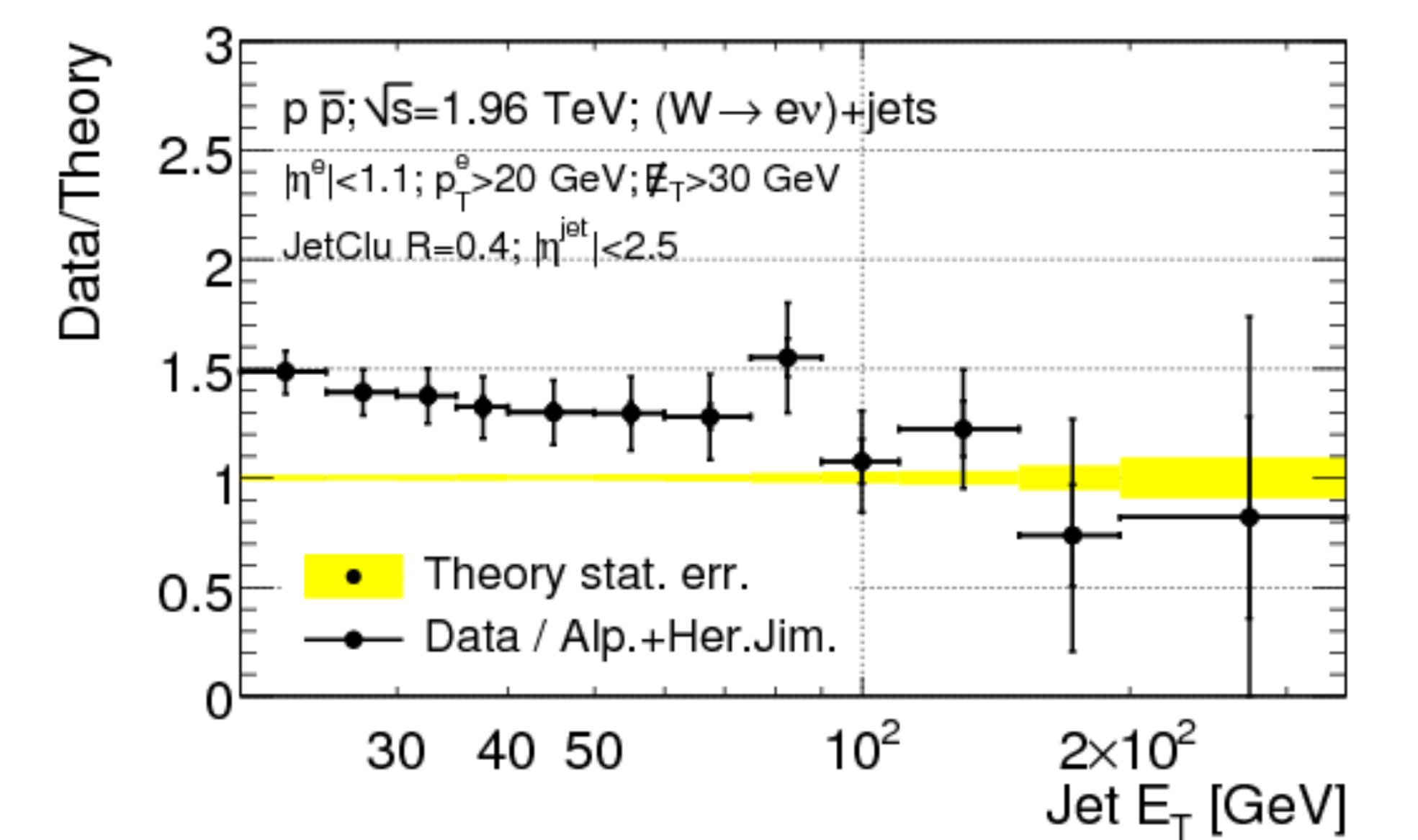}
\includegraphics[width=0.5\textwidth]{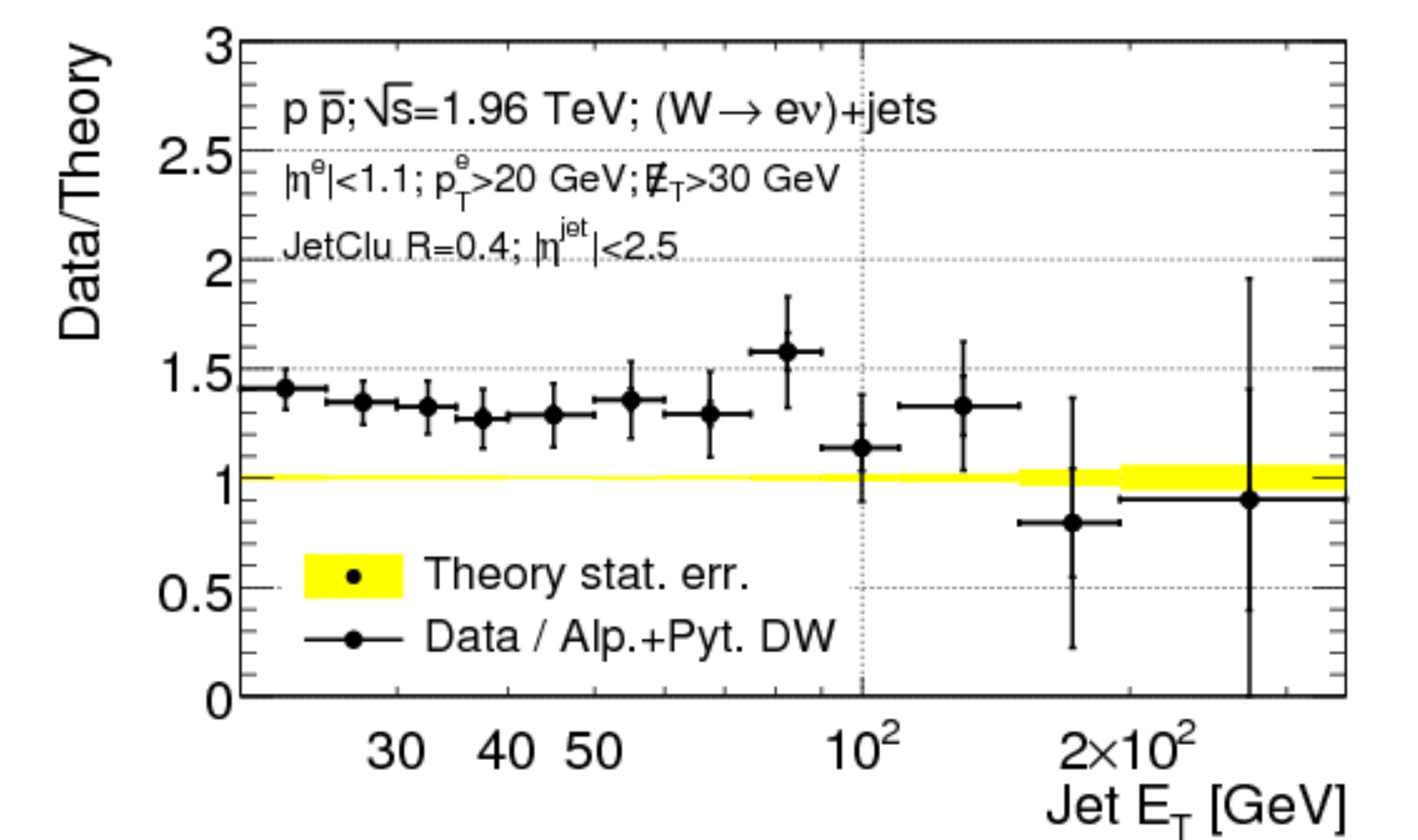}
\includegraphics[width=0.5\textwidth]{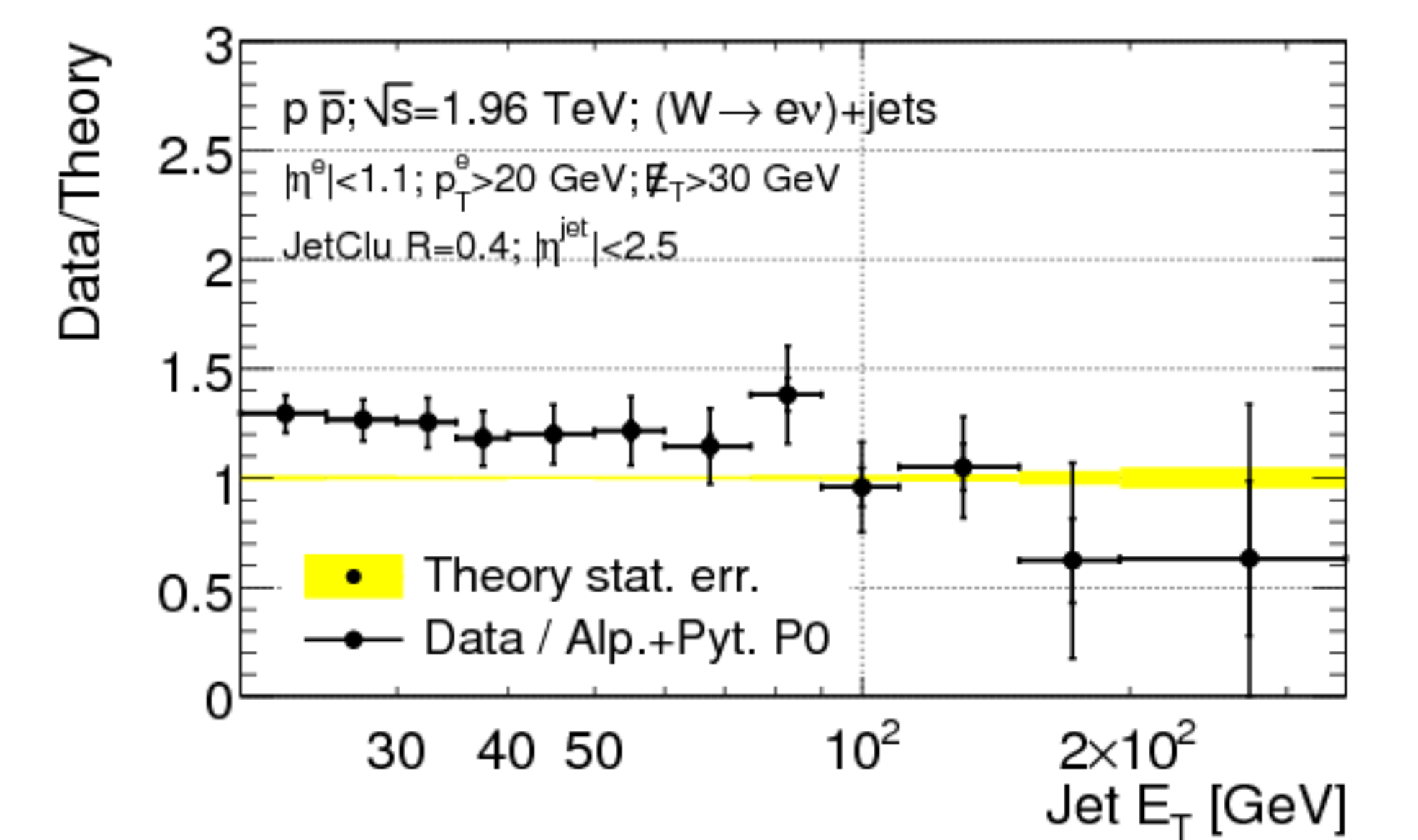}
\caption{Comparison of CDF data ~\cite{Aaltonen:2007ip} with the leading-jet $E_T$ spectrum
  predicted by \alpgen plus various MC codes and tunes.}
\label{fig:wjets_cdfvsMC}
\end{minipage}
\end{center}
\end{figure}

In Fig.~\ref{fig:wjets_MCcomp} we show the ratio of predictions for
the transverse energy ($E_T$) spectrum of the leading jet (that jet
with the highest $E_T$ per event) in $W$+jet final states at the
Tevatron, obtained by the merging of \alpgen with different shower
codes; \herwig, \pythiasix virtuality-ordered shower (DW), and $\pt$-ordered shower (\pzero).
 The differences between \herwig and \herwig plus \jimmy at small $E_T$ can  be explained
by the different amounts of energy that, in the various cases, are
deposited by the UE in the jet cones. In particular, as shown in
Fig.~\ref{fig:wjets_cdfvsMC}, these differences can accommodate the
slight shape discrepancy between data and \alpgen+\herwig that was
noted, at small $E_T$, in the CDF study~\cite{Aaltonen:2007ip}. It is
difficult, however, to attribute to the UE energy 
the significant differences seen in
Fig.~\ref{fig:wjets_MCcomp} at large $E_T$.

In order to investigate the source of the differences in the
predictions, systematic parameter variations of
the perturbative and non-perturbative model components of \pythiasix have
been studied using the Perugia family of \pythiasix tunes 
with \pzero as the central tune and
\phard  and \psoft  as the systematic variation tunes. The \psoft and \phard tunes both use the same Parton Density
Function (PDF) as \pzero, CTEQ5L~\cite{Lai:1999wy}, but
differ in the values of \pythiasix parameters controlling both perturbative and non-perturbative activity levels. In comparison to \psoft, the \phard tune has more perturbative
(initial and final state radiation) activity but less non-perturbative
(multiple interactions, beam remnant and hadronization) activity.
\psoft on the other hand has less perturbative but more
non-perturbative activity than the \pzero tune. In order to
investigate the interplay of the tuning variations with the MLM
matching, the effect of the change of the tune on the physics observables in both the \pythiasix standalone case and
\alpgenpythiasix case is presented.

\begin{figure}
\begin{center}
\begin{minipage}{150mm}
\includegraphics[width=0.5\textwidth]{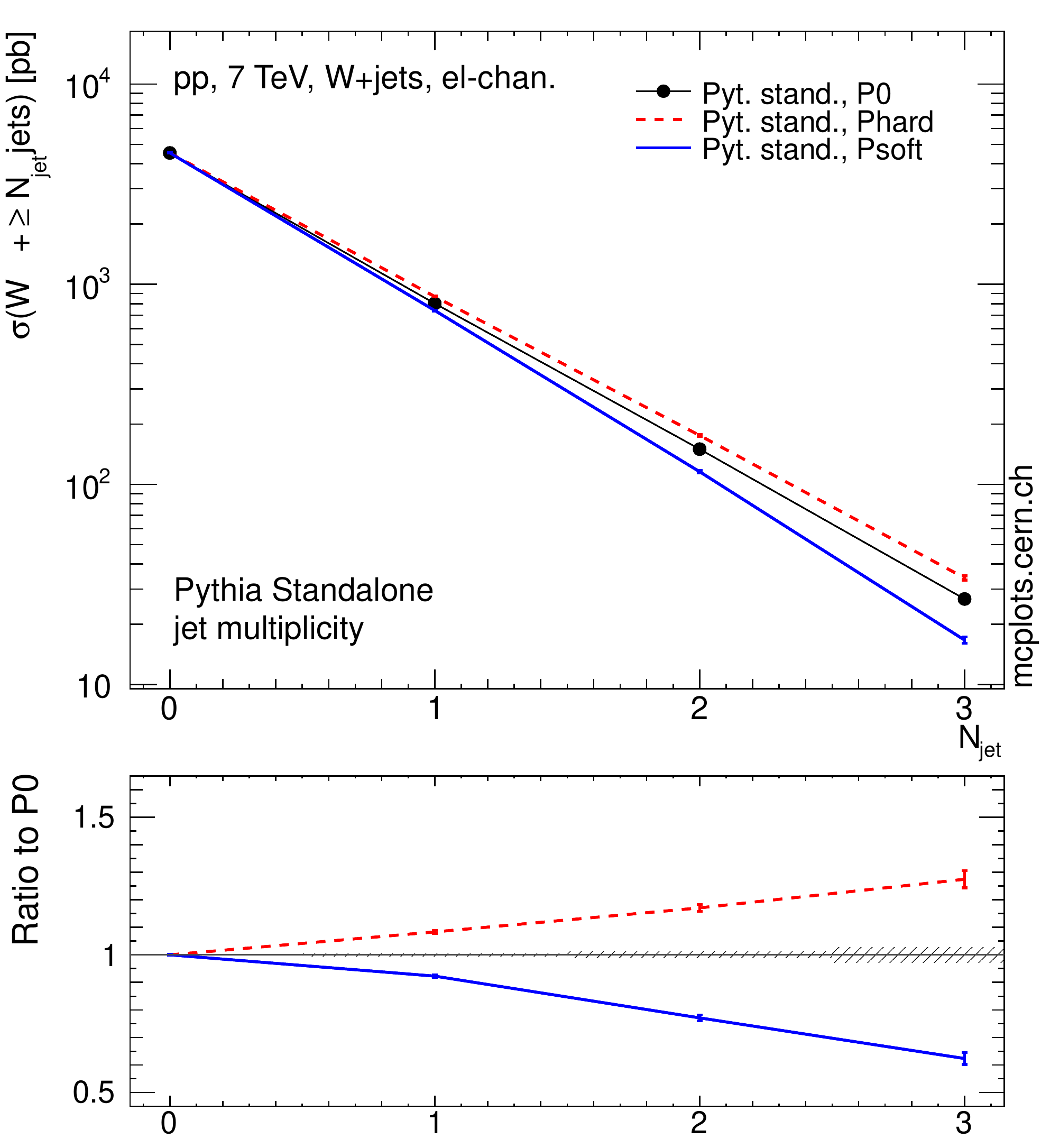}
\includegraphics[width=0.5\textwidth]{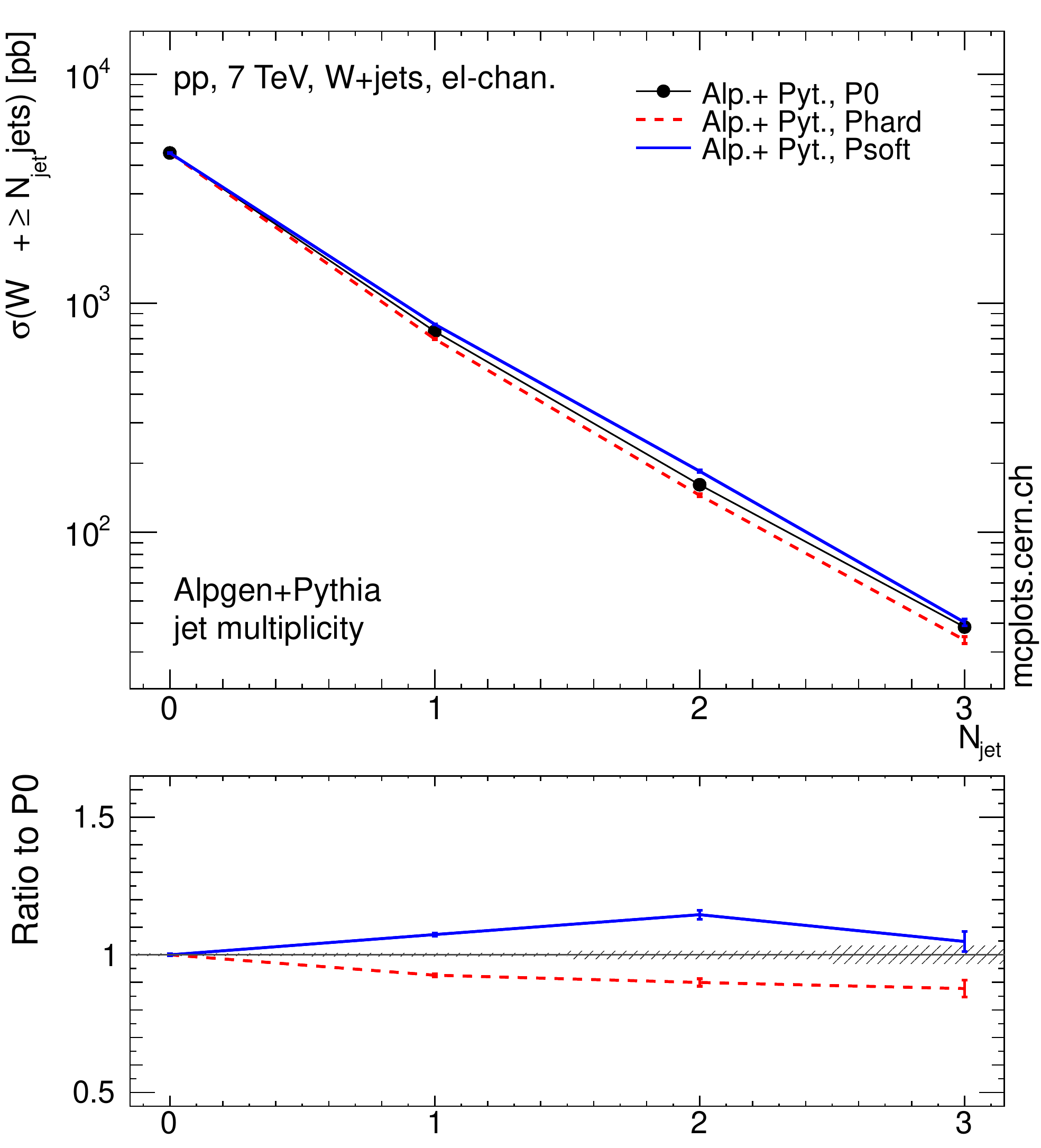}
\caption{Jet ($\pt>20~\rm{GeV}$ and $|\eta|<2.8$) multiplicity
  distribution in $W$+jets electron channel events in pp collisions at
  7 TeV. Distributions are shown for the samples generated with
  \pythiasixstand (left) and with \alpgenpythiasix (right). For each
  case the distributions obtained when using \pzero (P0), \phard (Phard) and \psoft (Psoft) 
  tunes are shown. All distributions are scaled so that the value of the first bin agrees with the ATLAS measurement~\cite{ATLAS-CONF-2011-060}.}
\label{fig:ATLAS_alpgen_stand_p0_wjet}
\end{minipage}
\end{center}
\end{figure}

In Fig.~\ref{fig:ATLAS_alpgen_stand_p0_wjet} the distribution of jet multiplicity
($\rm{N_{jet}}$) in $W$+jets events is compared for events generated with the \pzero, \phard and \psoft tunes. 
The $\rm{N_{jet}}$ observable is defined at the
particle-level according to the definition used in the ATLAS
measurement of the $W$+jets cross-section at $\sqrt s$=7
TeV~\cite{ATLAS-CONF-2011-060}. Jets are clustered from stable particles using the anti-Kt jet algorithm
\cite{Cacciari:2008gp} with the radius parameter $R = 0.4$, and considered in case they satisfy the following kinematic cuts: $\pt > 20~\rm{GeV}$ and $|\eta|<2.8$.
Comparisons are
performed for both the \pythiasixstand (left) and
\alpgenpythiasix (right)
cases.  For the \pythiasixstand case we observe that \phard tune yields
more high-$\pt$ jets than the \pzero tune while \psoft yields less final
state jets correspondingly. For the \alpgenpythiasix case an opposite trend is observed: 
\phard tune yields less high-$\pt$ jets
and \psoft tune yields more high-$\pt$ jets. In order to determine which modelling components of the \psoft and \phard tunes cause this behaviour,
we considered the effect of varying individual sets of parameters of
the \psoft and \phard tunes in \alpgenpythiasix predictions. Parameters were grouped
according to the modelling aspect they control into Initial State
Radiation (ISR), Final State Radiation including the FSR from the ISR
partons (FISR), the Underlying Event (UE) and Colour Reconnections (CR) blocks\footnote{The parameter blocks  organisation is similar to the one introduced in 
\cite{Skands:2010ak} and are listed in \ref{sec:tunes}}

Dedicated samples where only parameters of an
individual block were varied in the ranges used in \psoft and \phard
tunes on top of the \pzero tune were produced. The results of the
study, in terms of the cross-section contribution of each \alpgen
sub-sample (after MLM matching), are given in
Table~\ref{tab:P0variations}. 
As was already noted in Fig.~\ref{fig:ATLAS_alpgen_stand_p0_wjet}, the cross-section for
multijet production in the \phard case decreases with respect to
\pzero, and vice versa for \psoft. 
From Table~\ref{tab:P0variations}, we see that the parameter blocks
that produce this affect are the ISR and FISR blocks, while the
impact of the CR and UE block 
variations on the cross-sections is negligible. In addition to the simultaneous variations of parameters in the blocks, 
we have also performed individual parameter variations for each of the parameters in order to check that 
potential correlations between the parameters do not affect the
conclusions. Studies have also been performed for the Hadronization and Beam Remnant blocks of \cite{Skands:2010ak}. 
The variations of these parameters also had a negligible effect on the kinematic distributions and cross-section values.

\begin{figure}
\begin{center}
\begin{minipage}{150mm}
\includegraphics[width=0.5\textwidth,angle=0]{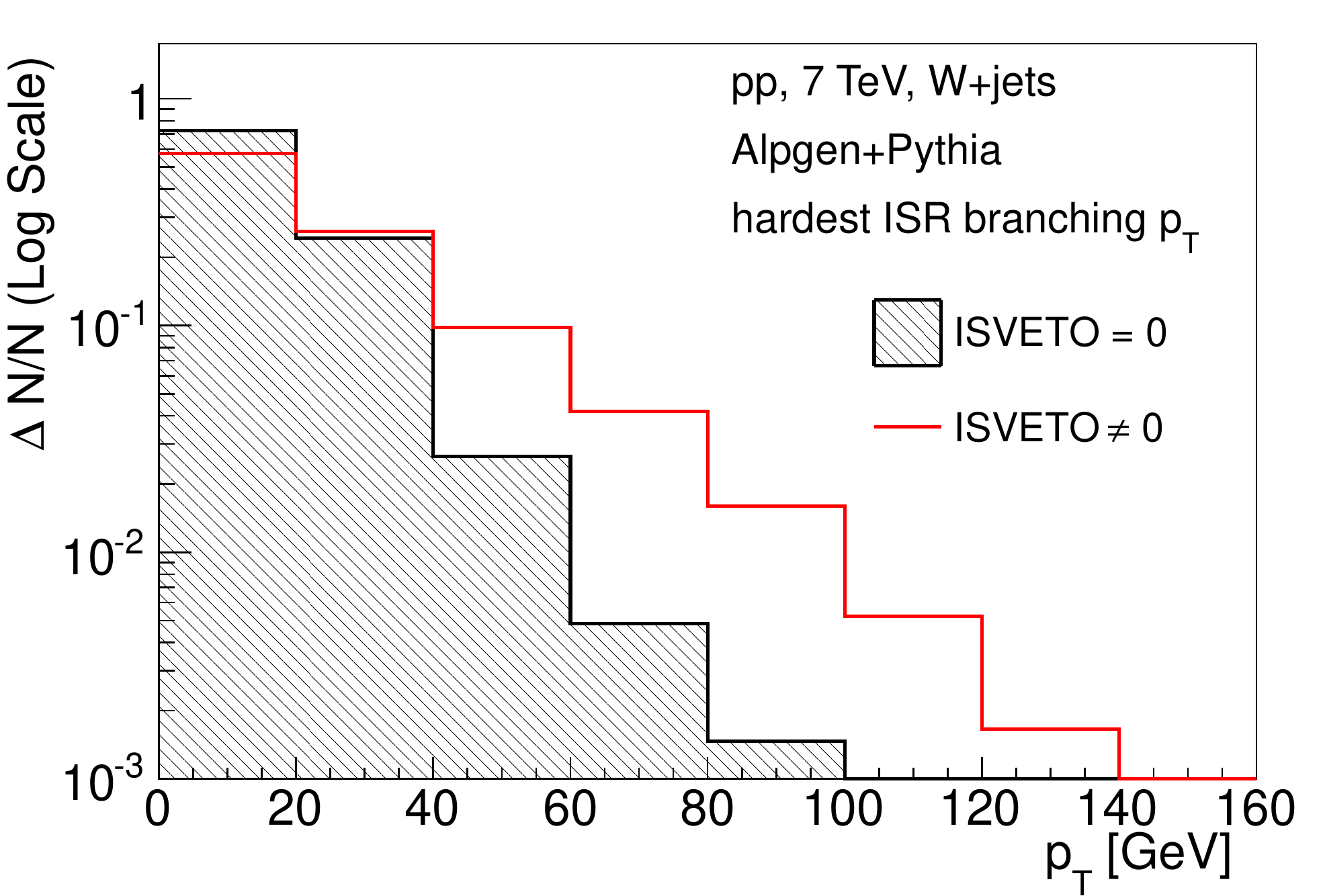}
\includegraphics[width=0.5\textwidth,angle=0]{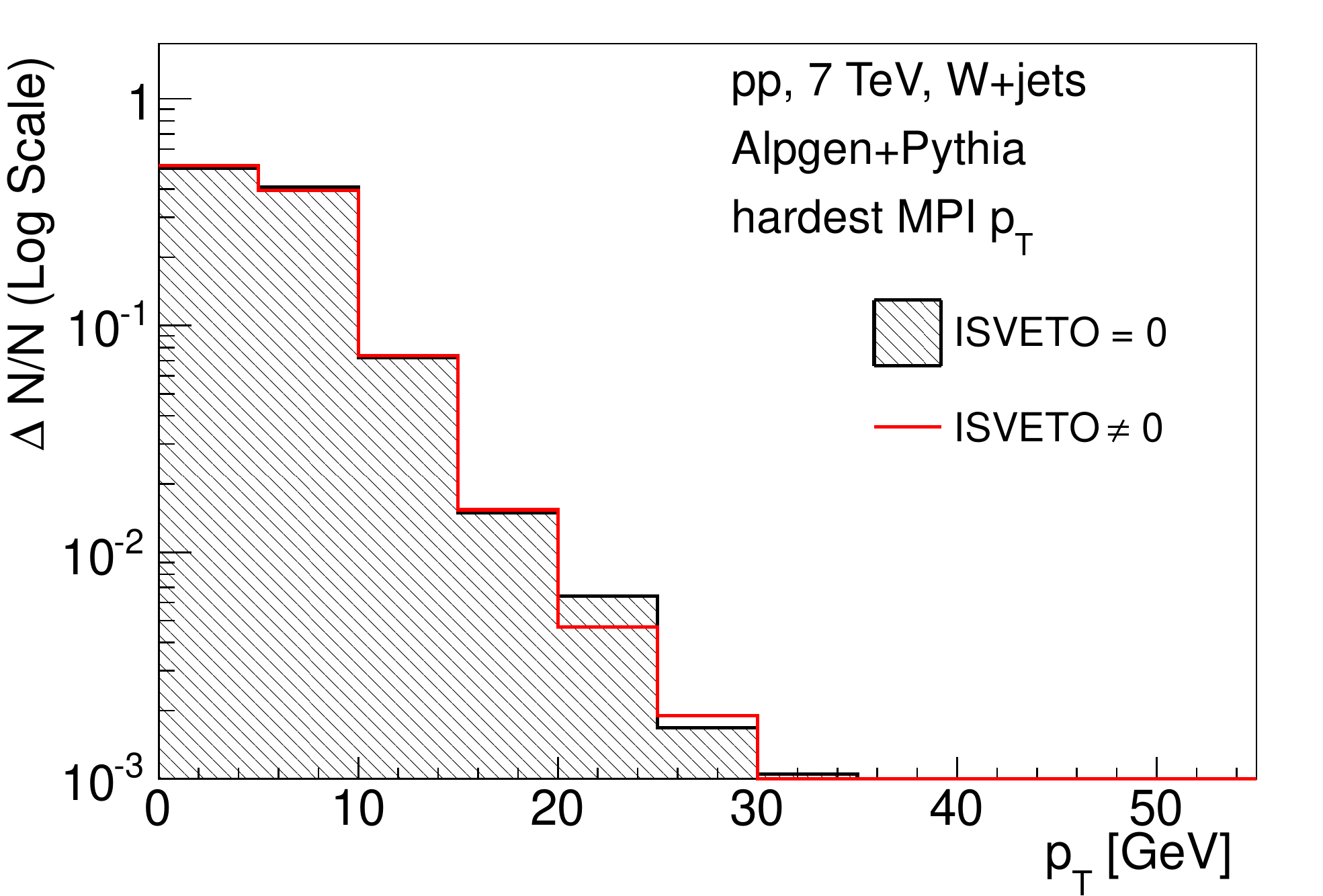}
\caption{ Distribution of the probabilities for the event acceptance
  (ISVETO=0) or rejection (ISVETO$\neq$0) during the MLM matching
  step, as a function of the largest $\pt$ shower emission from the
  initial state radiation (left) and the largest $\pt$ multiple
  proton-proton interaction in the event (right). The events were
  generated using exclusive sub-sample of \alpgenpythiasix \pzehn
  $W$+jets events with exactly three additional partons from the
  matrix element in the final state for pp collisions at 7 TeV and
  \alpgenpythiasix \pzehn tune.}
\label{fig:rejection_vs_kt}
\end{minipage}
\end{center}
\end{figure}

In Fig.~\ref{fig:rejection_vs_kt} we demonstrate that the increased
parton shower activity can indeed lead to the reduced cross-section
(and softer jet spectra) due to the increased rates at which the
\alpgenpythiasix events are vetoed during the MLM matching. In the
figure the distributions of the events that pass or fail (ISVETO=0 or
ISVETO$\neq$0) the MLM matching criterion are shown for the exclusive
sub-sample of \alpgenpythiasix \pzehn $W$+jets events with exactly
three additional partons from the matrix element in the final
state\footnote{The observations in the text are largely independent on
  the final state parton multiplicity}. Each of the distributions is
normalised to unit area. The distributions are shown as a function of
the largest $\pt$ shower emission from the initial state radiation
(left) and as a function of the largest $\pt$ multiple proton-proton
interaction. \footnote{These $\pt$ values are reported by \pythiasix
  parameters \ttt{VINT(357)} (ISR) and \ttt{VINT(359)} (MPI)
  respectively.} In the left hand side figure we see that the events
are rejected with higher probability, the larger the $\pt$ of the
hardest ISR branching in the event. Therefore, a \pythiasixstand tune
which increases the ISR activity can, somewhat counter-intuitively,
reduce the rate for multijet and hard emissions. In the right hand
side we demonstrate that the events are accepted and rejected
independently of the transverse momentum of the hardest multiple
interaction in the event (which is the desired behaviour of the
matching application used with the parton shower code).

\begin{table}
\begin{center}
\footnotesize{
\begin{tabular}{lllllll}
\hline
tune & Np0 & Np1 & Np2 & Np3 & Np4+ & total [pb] \\
\hline
Phard &  7287 $\pm$  3.9 &  728 $\pm$  2.6 &  141 $\pm$  1.3 &   27 $\pm$  0.2 &   6.6 $\pm$  0.2 &  8190 $\pm$  8\\
P0 &   7556$\pm$  3.6 &  814 $\pm$ 2.7 &  166 $\pm$  1.3 &   32 $\pm$ 0.3 & 7.8  $\pm$ 0.3 &  8576 $\pm$ 8 \\
Psoft    &  7804 $\pm$  3.4 &  944 $\pm$  2.8 &  207 $\pm$  1.5 &   42 $\pm$  0.3 &   10.1$\pm$  0.3 &  9007 $\pm$  8\\
\hline
P0 with Phard ISR  &   7207 $\pm$  6.9 &   735 $\pm$  2.6 &  143 $\pm$  1.3 &  27 $\pm$  0.2 &  6.9 $\pm$  0.2 &  8119 $\pm$ 11 \\
P0 with Psoft ISR   &  7831 $\pm$  4.9 &   881 $\pm$  2.7 &  186 $\pm$  1.4 &  36 $\pm$  0.3 &  8.8 $\pm$  0.3 &  8943 $\pm$ 10 \\
\hline
P0 with Phard FISR   & 7548 $\pm$  6.0&    814 $\pm$  2.7 &  167 $\pm$  1.3 &  32 $\pm$  0.3 &  7.8 $\pm$  0.3 &  8569 $\pm$ 10 \\
P0 with Psoft FISR  &  7505 $\pm$  6.1&   878 $\pm$  2.7 &   188 $\pm$  1.4 &  37 $\pm$  0.3 &  9.4 $\pm$  0.3 &  8617 $\pm$ 10 \\
\hline
P0 with Phard UE   &   7513 $\pm$  6.1 &   826 $\pm$  2.7 &  171  $\pm$  1.4 &  33 $\pm$ 0.3 &  7.8 $\pm$  0.3 &  8551 $\pm$ 10 \\
P0 with Psoft UE  &   7576 $\pm$  5.9 &   817$\pm$  2.7 &  166 $\pm$ 1.3 &   32 $\pm$ 0.3 &   8.1 $\pm$  0.3 &   8599 $\pm$ 10 \\
P0 with Phard CR   &  7561 $\pm$  5.9 &   821 $\pm$  2.7 & 167 $\pm$  1.3 & 32  $\pm$ 0.3  &  8.1 $\pm$  0.3 &   8589 $\pm$ 10 \\
P0 with Psoft CR  &   7556 $\pm$  5.9 &   815 $\pm$  2.7 & 165  $\pm$ 1.3 & 32  $\pm$ 0.3&   8.1 $\pm$  0.3 &   8576 $\pm$ 10 \\
\hline
\end{tabular}
}
\caption{Impact of different variations of the \pythiasix \pzehn tunes on the cross
  sections of \alpgen $W$+jets sub-samples with different matrix element parton multiplicities, and the total
  inclusive $W$ cross section. For the studies sub-samples with up to four additional partons from the 
matrix element were generated. The matching is performed inclusively for the highest parton multiplicity sub-sample 
and exclusively for other sub-samples. The tabulated cross-sections were extracted after the MLM
  matching and parton shower. The errors shown are statistical only. The parameter settings of the various
  setups are discussed in the text.
\label{tab:P0variations}
}
\end{center}
\end{table}

To conclude, the origin of the differences observed in the predictions
of tunes with different ISR/FSR activity matched to \alpgen is rather
due to the mismatch between the jet-emission probability predicted by
the matrix elements and by the shower. This comes from the mismatch in
the value of \alphaS discussed earlier, arising from different values
of \LambdaQCD or from the use of a different evolution variable in the
shower.  If the value of \alphaS in the shower increases, the emission
rate of additional jets during the shower evolution will
increase. Since the matching algorithm rejects events with extra jets
generated by the shower, to replace them with events where the jet is
accounted for by a higher-order matrix element calculation, a larger
value of \alphaS in the shower leads to a higher rejection
rate. Unless this change in \alphaS is accompanied by a similar change
in the matrix element calculation, the additional rejection is not
compensated by the relative increase in rate for the higher-order
parton-level contributions, leading to the effects reported in this
section.

This important interplay between MC parameters, which are typically
tuned to ``soft'' observables such as UE or the small-$\pt$ DY
spectrum, and the performance of the matching algorithms for ``hard''
observables, calls for particular attention when adopting new UE tunes
in the framework of multijet studies with matrix-element
matching. Along the same lines, it should be kept in mind that, tuning
a stand-alone shower MC to better model multijet final states, will
force it to emulate effects present in the multiparton matrix
elements. Using such tunes with matrix-element matching therefore
requires ad-hoc modifications of the matching algorithm, or of its
parameters.

\section{Stabilising ME-PS Matched Tunings\label{sec:stabalising}}

In this section we discuss how to overcome the problems discussed in
the previous section with a simple prescription, and outline a tuning
strategy that should allow to consistently optimize, in the context of
the \pythiasix shower MC, the description of both the UE and the
high-$E_T$ properties of final states.

\subsection{A New \alpgenpythiasix \alphaS~Consistent Tune\label{sec:newtune}}

As it was explained in Section~\ref{sec:theory}, and practically
demonstrated in Section~\ref{sec:example}, it is highly desirable to
have a consistent treatment of \alphaS on either side of the ME and PS
boundary.  In Appendix~\ref{sec:aptune} the relevant settings for a
new \alphaS~consistent \alpgenpythiasix tune are described in detail.
In this tune, the \alphaS~consistency is essentially ensured by
setting the effective value of \lamqcd~to be the same throughout the
\pythiasix~parton shower algorithms and in the \alpgen matrix
elements. A consistent choice for \LambdaQCD of
\begin{equation}
\Lambda^{(5)}_{\mrm{QCD}} \sim 0.26~,
\end{equation}
is made, where the superscript indicates the number of flavours. This choice is
informed by comprehensive \professor tunings  
\cite{Buckley:2009bj, AtlasTuneNote} of the $\pt$-ordered shower in \pythiasix
\cite{Sjostrand:2004ef} to event shapes and other LEP data. Note that the settings for \pythiasix~are those of the central
\pelf (P2011) tune~\cite{Skands:2010ak}, which was inspired by these
studies. We will refer to this new tune of \alpgenpythiasix as the
\pelfmatched tune.

\subsection{Tests of the Consistent \alphaS~Approach: Behaviour Under
  Scale Variations}

\begin{figure}
\begin{center}
\begin{minipage}{150mm}
\includegraphics[width=0.5\textwidth]{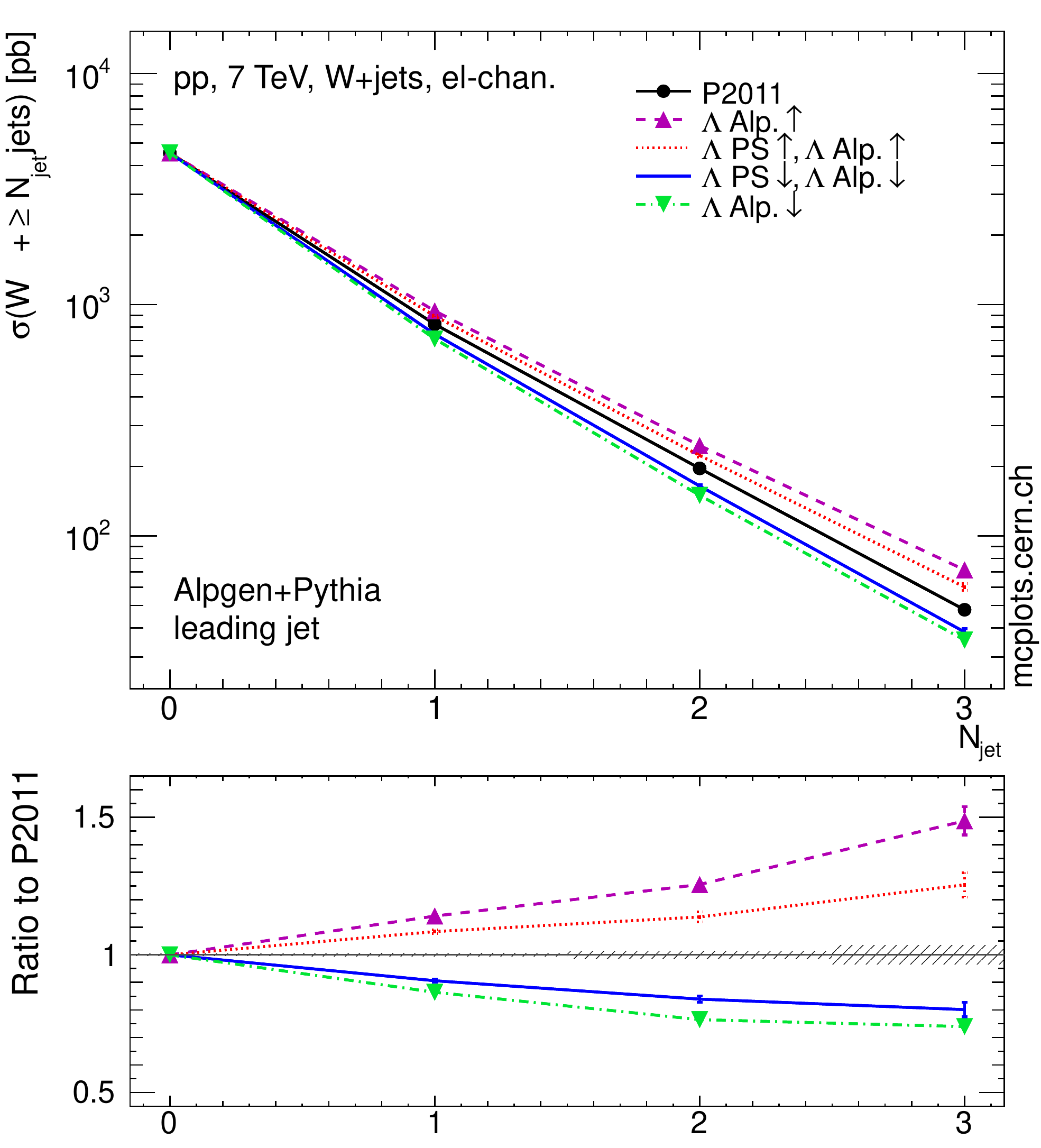}
\includegraphics[width=0.5\textwidth]{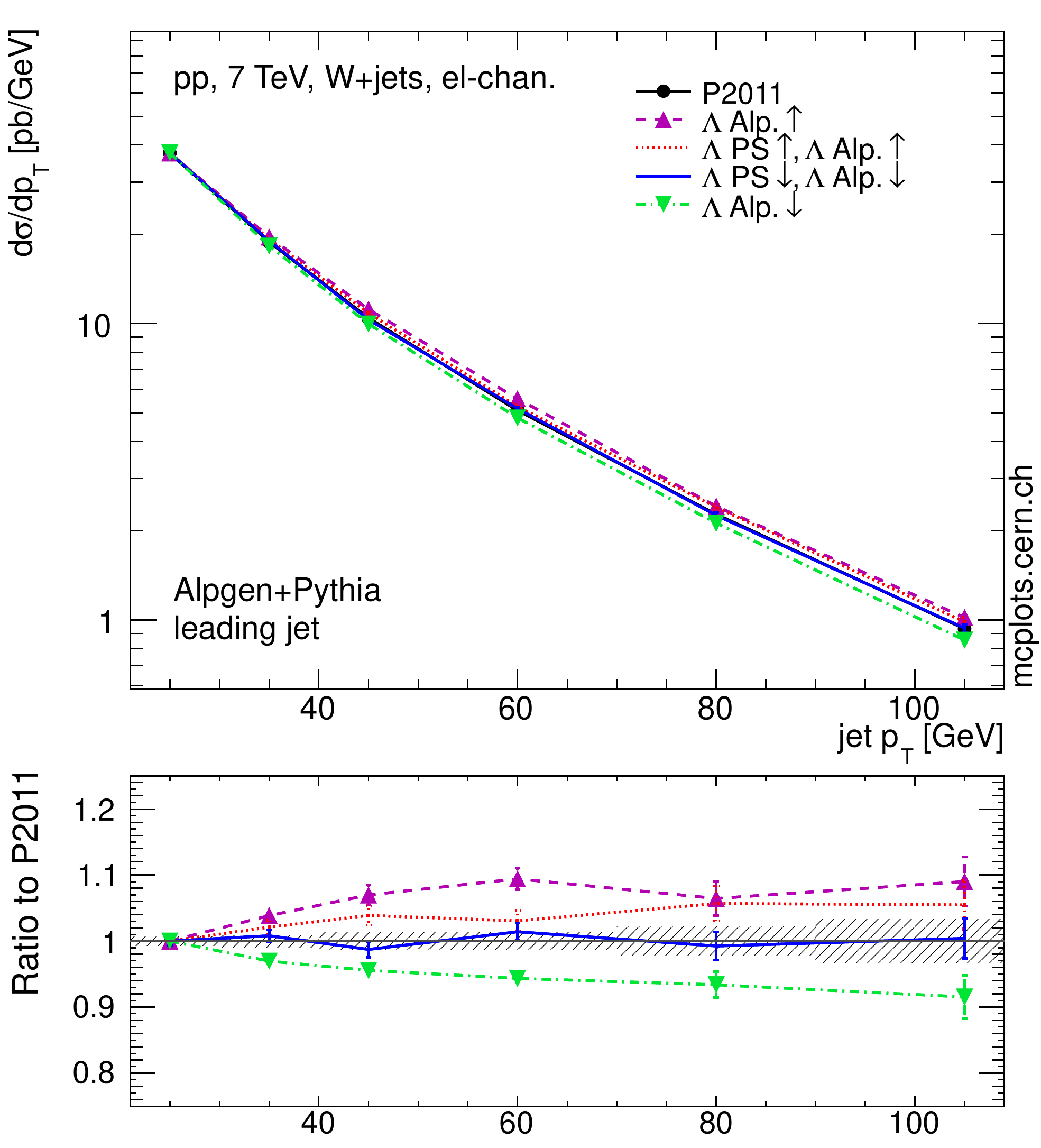}
\caption{Comparison of \alpgenpythiasix ($\pt>$20~GeV) jet
  multiplicity (left) and leading jet transverse momentum (right)
  distributions in $W$+jets electron channel events. The samples are
  generated using different \alpgenpythiasix parameter setups described in the text.}
\label{fig:ATLAS_alpgen_stand_p0_wjet_stab}
\end{minipage}
\end{center}
\end{figure}

In this section we study the behaviour of the new \alpgenpythiasix \pelfmatched tune
under \lamqcd~variations to demonstrate that, with a consistent treatment of
\alphaS, the expected behaviour of ME-PS matched predictions under variations
of tuning parameters is restored. $W$+jets events
selected with the same criteria applied for
Fig.~\ref{fig:ATLAS_alpgen_stand_p0_wjet} are used.
Figure~\ref{fig:ATLAS_alpgen_stand_p0_wjet_stab} shows the jet multiplicity 
(left) and leading jet transverse momentum (right) distributions for
the \pelfmatched tune and 
four variant tune samples generated with different \LambdaQCD~ values. Two samples,
labelled as  ``$\Lambda\mbox{ Alp. }\uparrow$'' and ``$\Lambda\mbox{ Alp. }\downarrow$'',
have \LambdaQCD~ respectively increased and decreased by a factor of 2 only in
the ME calculation. This is achieved by setting respectively the \alpgen~ parameter \ktfac~
 to 1/2 and 2. The increase (decrease) of the \LambdaQCD~ value 
in \alpgen~ results in more (less) jets and a harder (softer) leading jet
spectrum as shown in  Fig.~\ref{fig:ATLAS_alpgen_stand_p0_wjet_stab}.
The two samples labelled as ``$\Lambda\mbox{ PS }\uparrow,\ \Lambda\mbox{
  Alp. }\uparrow$'' and ``$\Lambda\mbox{ PS }\downarrow,\ \Lambda\mbox{
  Alp. }\downarrow$'' correspond to a consistent variation of \lamqcd~ both in 
the ME and PS, with \lamqcd~ respectively increased and decreased by a factor of 2.
The impact of these variations is qualitatively similar to the case where
\lamqcd~ is only varied in the ME, restoring the expected behaviour of
ME-PS matched prediction under variation of \lamqcd.
However, the samples with \lamqcd~ varied
simultaneously in the ME and in the PS exhibit a smaller deviation from the
nominal sample. The mitigation of the 
impact of a \lamqcd~ coherent change in a ME-PS matched sample compared to the same
change only in the ME calculation is due to the interplay between the radiation
produced by PS and the matching algorithm, as detailed in Section~\ref{sec:theory}. 
While the choice of the {\tt xlclu} parameter allows to directly adapt 
\alpgen\ to possible future changes in the choice of \lamqcd\ in \pythia, 
the variation of the \ktfac\ parameter in the standard range $0.5<\ktfac<2$ 
can be used to establish the range of the systematical uncertainty, or to 
tune the description of specific observables.

\section{Comparisons with Data\label{sec:datacomp}}

In this section we demonstrate that the new \LambdaQCD-consistent
\pelfmatched tuning of
\alpgenpythiasix introduced in Section~\ref{sec:newtune} compares well with recent
Tevatron and LHC measurements, and that, with the arrival of improved
precision measurements, there should be room for further tuning of
these predictions.

\subsection{$Z$/$W$+jets production\label{sec:wzjetsdata}}

The figures that follow show comparisons of \alpgenpythiasix Monte
Carlo predictions to measurements of published $Z$+jets and $W$+jets
processes from CDF~\cite{cdf10394,Aaltonen:2007ip,:2007cp} and
$W$+jets from ATLAS~\cite{Aad:2010pg}\footnote{Measurements of these
  processes at the Tevatron have also been performed by
  D0~\cite{Abazov:2008ez,Abazov:2009av,Abazov:2011rf}. Preliminary,
  higher statistics, studies of $W/Z$+jets production at the LHC have
  been reported by
  ATLAS~\cite{ATLAS-CONF-2011-060,ATLAS-CONF-2011-042}, and
  CMS~\cite{CMS-PAS-EWK-10-012}.}.  These cross-section measurements are
  corrected for all known detector effects to particle level and
  compared to Monte Carlo predictions.  The Monte Carlo predictions of
  $V$+jet production cross-sections are formed by clustering the
  stable final state particles ($\tau>10$~ps) following parton shower
  and hadronization of the unweighted events. This clustering is done
  using the same jet algorithm as the measurement, as implemented in
  the Fastjet \cite{Cacciari:2005hq} package. All stable final state
  particles are used, with the exception of the leptons that result
  from the decay of the signal $W$ or $Z$ boson. The decay leptons are
  corrected for the final state QED radiation such that their
  4-momentum is equivalent to that before radiation. After the events
  have been clustered, the restrictions on the allowed phase space of
  the jets and of the $W$/$Z$ boson decay products are applied to be
  consistent with the measurement to which we are comparing.  The
  prediction of the final \alpgenpythiasix cross-sections contains
  contributions from $V$+0, 1, 2, 3 parton samples (showered with
  exclusive MLM matching), and $V$+4 parton samples (showered with
  inclusive MLM matching).

Two different \alpgenpythiasix generations are compared to the
data; the new \alpgenpythiasix \pelfmatched tune introduced in
Section~\ref{sec:newtune} (labelled ``Alp.+Pyt. P2011''), and an
\alpgenpythiasix prediction using the default settings of \alpgen and the
\pythiasix DW tune (labelled ``Alp.+Pyt. DW''). The ratio of the
matched predictions to the data are shown and compared. 
Additionally, the results of variations of \ktfac by factors of 0.5 and 2.0 in the
\alpgenpythiasix \pelfmatched prediction are shown as solid lines. The
hatched regions show the total error (statistical plus systematic)
propagated to the theory/data ratio from the data measurements. The
error bars on the points show the statistical error on the theoretical
prediction.

\begin{figure}
\begin{center}
\begin{minipage}{150mm}
\includegraphics[width=0.5\textwidth]{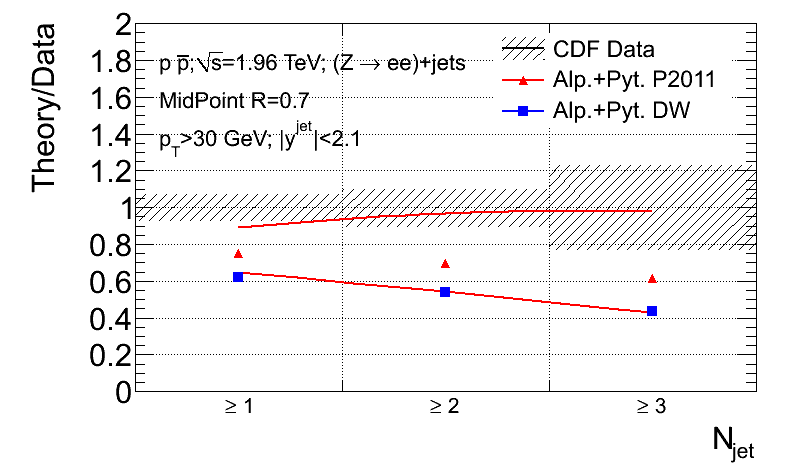}
\includegraphics[width=0.5\textwidth]{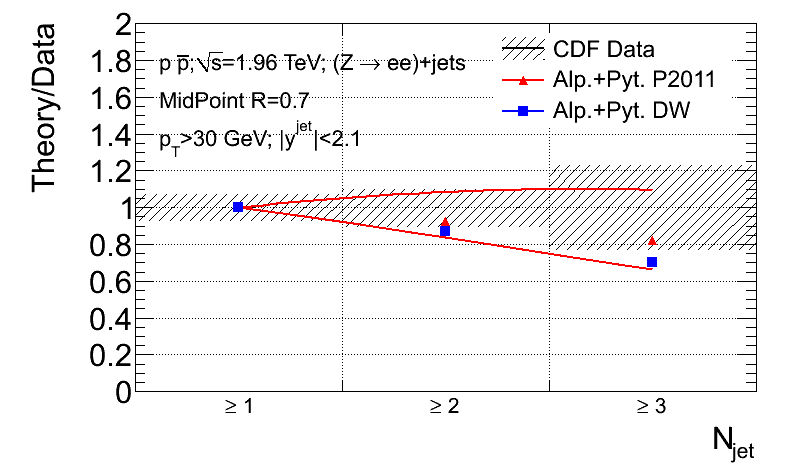}
\includegraphics[width=0.5\textwidth]{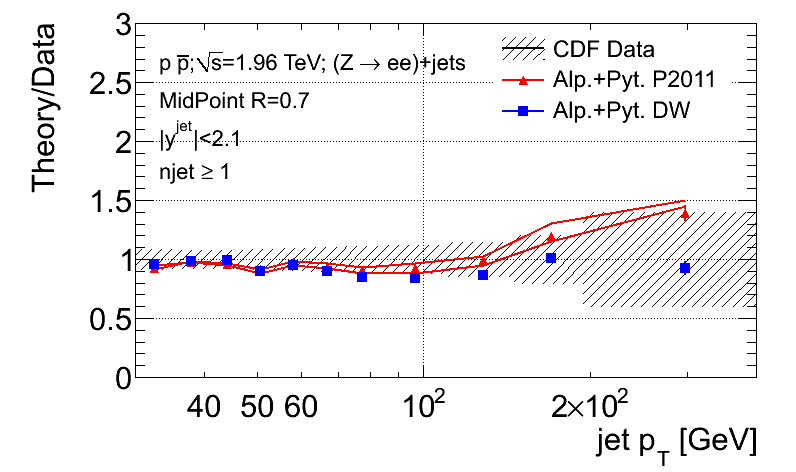}
\includegraphics[width=0.5\textwidth]{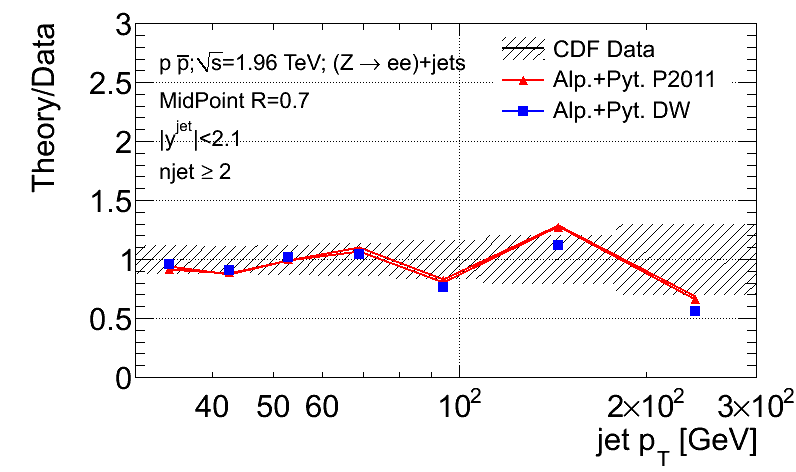}
\caption{ (Top) The ratio of predicted theory and CDF measured data
  cross-sections for the production of a $Z\rightarrow ee$ boson in association with
  at least $\rm{N_{jet}}$ jets \cite{:2007cp}. In the
left hand figure the theory predictions are not normalised to the
data. In the right hand figure the theory predictions are normalised
such that they equal the data measurement in the $\geq 1$ jet bin.
(Bottom) The ratio of predicted theory and CDF measured data
  cross-sections for the production of a $Z\rightarrow ee$ boson in association with
  at least 1 jets (left hand side) and at least 2 jets (right hand side) as a function of jet
  $E_{T}$. In the left hand plot, the theory prediction is normalised
  such that the predicted rate for $\geq 1$ jet production is equal to
  that measured in the data. In the right hand plot, the theory prediction is normalised
  such that the predicted rate for $\geq 2$ jet production is equal to
  that measured in the data.}
\label{fig:cdfz_njet}
\end{minipage}
\end{center}
\end{figure}

\begin{figure}
\begin{center}
\begin{minipage}{150mm}
\includegraphics[width=0.5\textwidth]{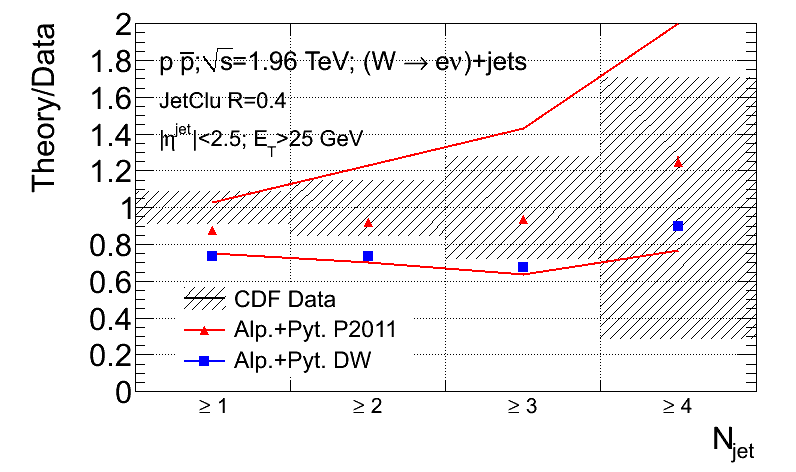}
\includegraphics[width=0.5\textwidth]{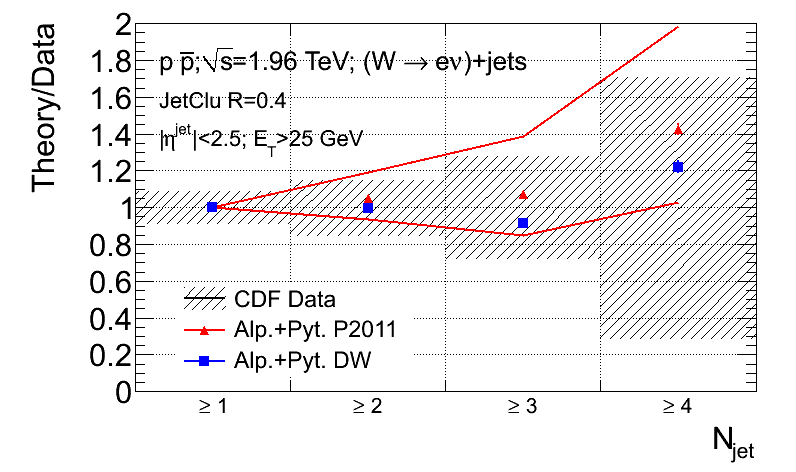}
\includegraphics[width=0.5\textwidth]{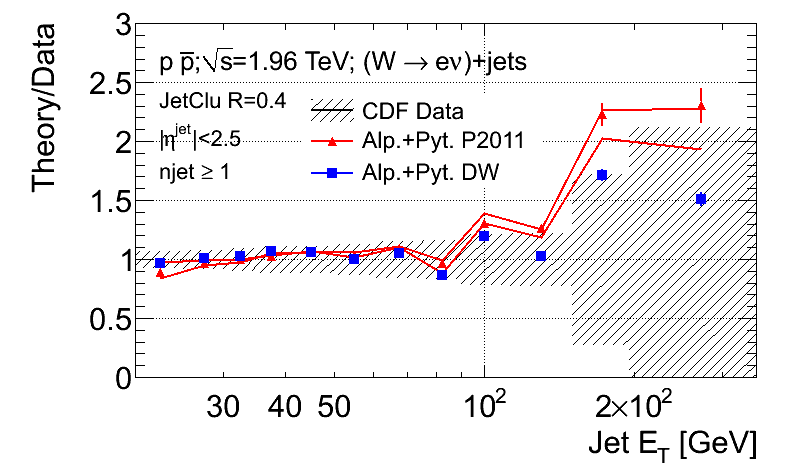}
\includegraphics[width=0.5\textwidth]{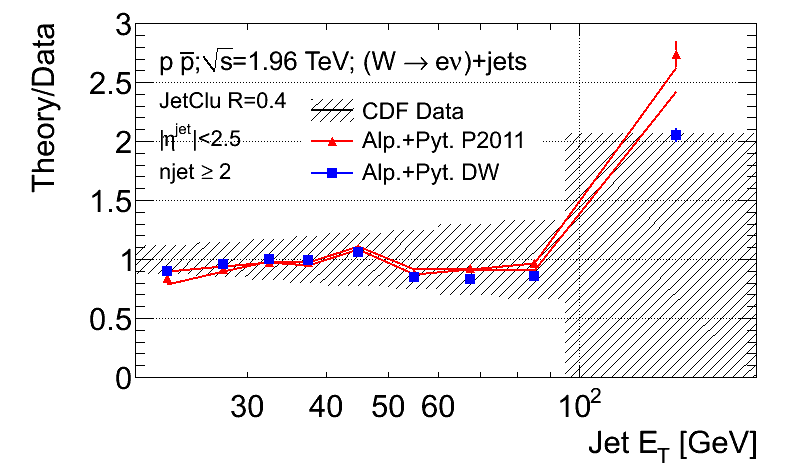}
\caption{(Top) The ratio of predicted theory and CDF measured data 
  cross-sections for the production of a $W\rightarrow e\nu$ boson in association with
  at least $\rm{N_{jet}}$ jets \cite{Aaltonen:2007ip} . In the
left hand figure the theory predictions are not normalised to the
data. In the right hand figure the theory predictions are normalised
such that they equal the data measurement in the $\geq 1$ jet bin.
(Bottom) The ratio of predicted theory and CDF measured data
  cross-sections for the production of events containing a $W\rightarrow e\nu$ boson in association with
  at least 1 jets (left hand side) and at least 2 jets (right hand
  side), as a function of the leading jet
  $E_{T}$ (left hand side), and the sub-leading jet $E_{T}$ (right
  hand side). In the left hand plot, the theory prediction is normalised
  such that the predicted rate for $\geq 1$ jet production is equal to
  that measured in the data. In the right hand plot, the theory prediction is normalised
  such that the predicted rate for $\geq 2$ jet production is equal to
  that measured in the data.
}
\label{fig:cdfw_njet}
\end{minipage}
\end{center}
\end{figure}

\begin{figure}
\begin{center}
\begin{minipage}{150mm}
\includegraphics[width=0.5\textwidth]{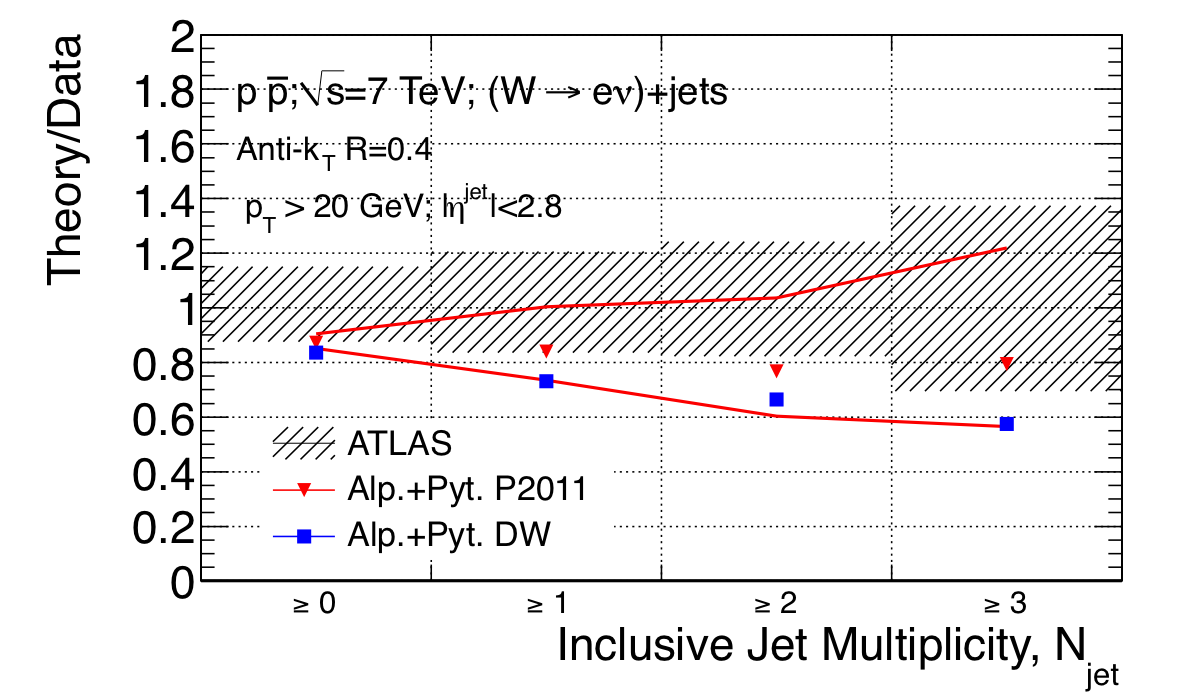}
\includegraphics[width=0.5\textwidth]{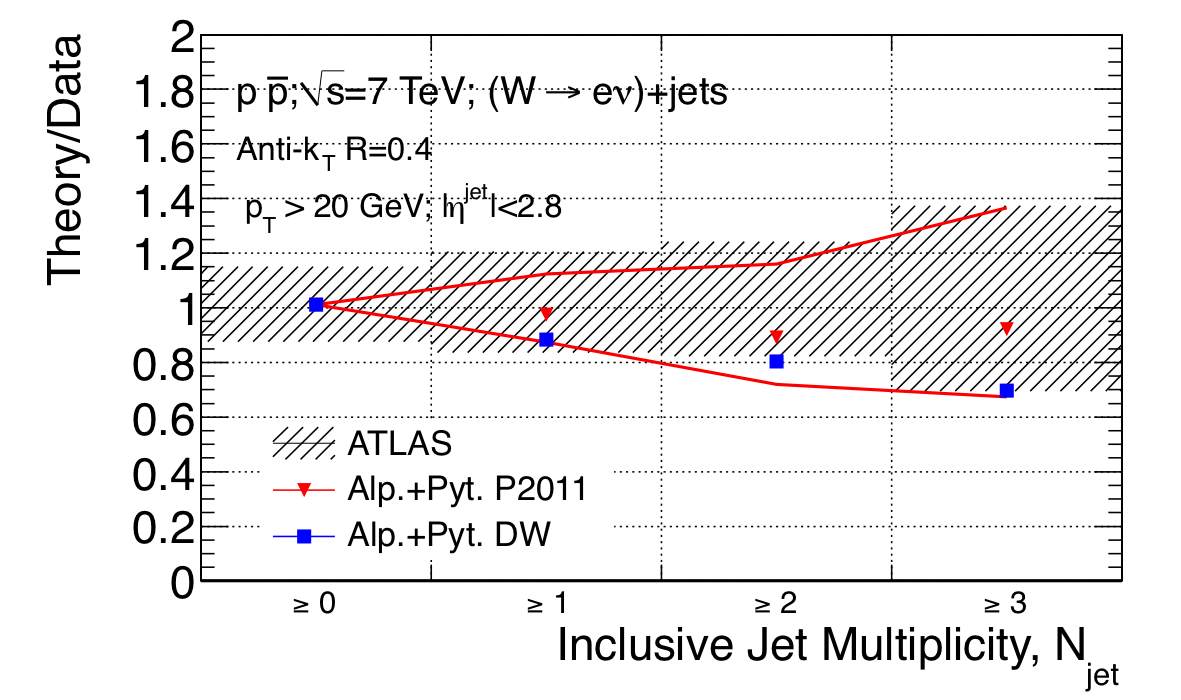}
\includegraphics[width=0.5\textwidth]{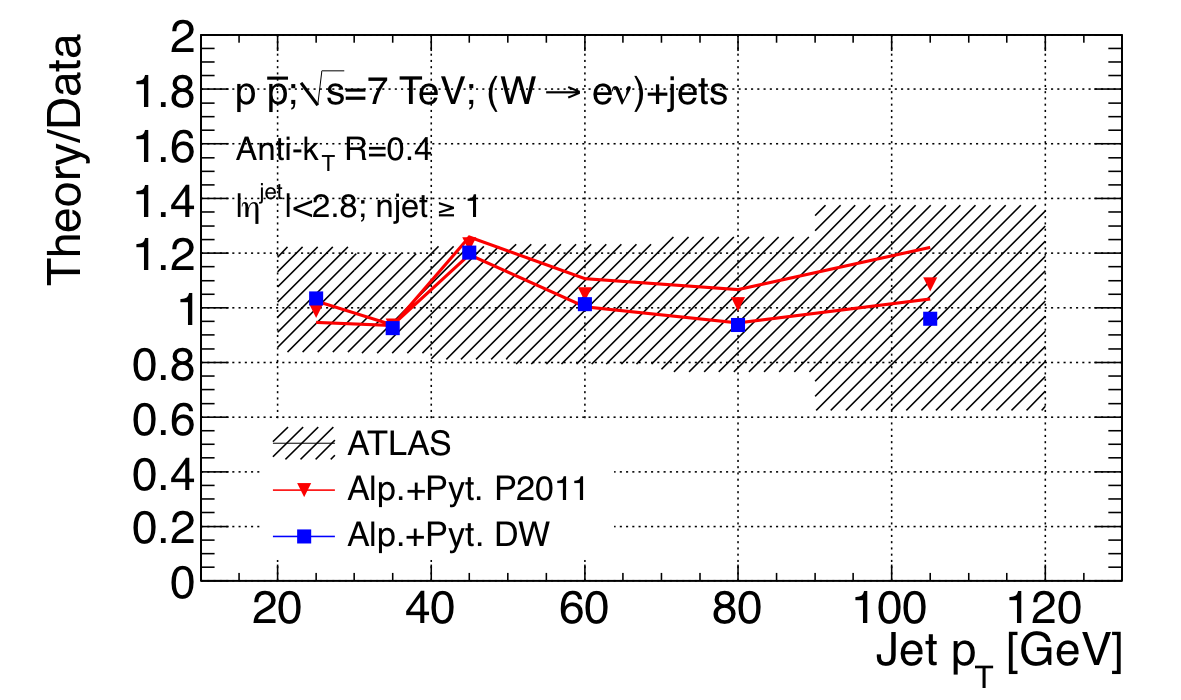}
\includegraphics[width=0.5\textwidth]{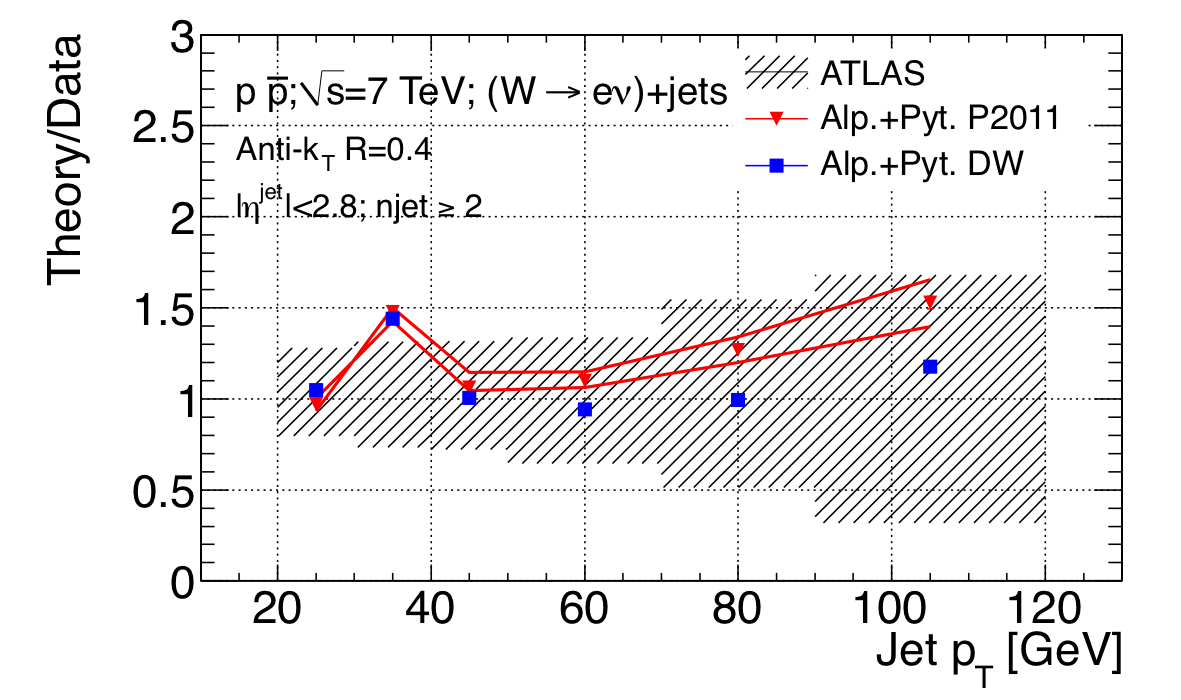}
\caption{(Top) The ratio of predicted theory and ATLAS published
  cross-sections~\cite{Aad:2010pg} 
for the production of a $W\rightarrow e\nu$ boson in association with
  at least $\rm{N_{jet}}$ jets. In the
left hand figure the theory predictions are not normalised to the
data. In the right hand figure the theory predictions are normalised
such that they equal the data measurement in the $\geq 0$ jet bin.
(Bottom) The ratio of predicted theory and ATLAS data
  cross-sections for the production of events containing a $W\rightarrow e\nu$ boson in association with
  at least 1 jets (left hand side) and at least 2 jets (right hand
  side), as a function of the leading jet
  $\pt$ (left hand side), and the sub-leading jet $\pt$ (right
  hand side). In the left hand plot, the theory prediction is normalised
  such that the predicted rate for $\geq 1$ jet production is equal to
  that measured in the data. In the right hand plot, the theory prediction is normalised
  such that the predicted rate for $\geq 2$ jet production is equal to
  that measured in the data.}
\label{fig:atlaswele_njet}
\end{minipage}
\end{center}
\end{figure}

In Figure~\ref{fig:cdfz_njet} we show the ratio of the predicted
theory cross-sections to the data for the CDF $Z$+jets measurement
\cite{:2007cp}. In this measurement jets are defined by the CDF midpoint
algorithm \cite{Abulencia:2005yg}, with $R_{cone} = 0.7$
and are required to have $p^{jet}_{T} > 30~\rm{GeV}$ and $|y^{jet}| <
2.1$. 
In Figure~\ref{fig:cdfw_njet} we show the ratio of the predicted
theory cross-sections to the data for the CDF $W$+jets measurement
\cite{Aaltonen:2007ip}. In this measurement jets are defined by the CDF JetClu
algorithm \cite{Abe:1991ui}, with $R_{cone} = 0.4$
and are required to have $\pt > 20~\rm{GeV}$ and $|\eta| <
2.5$. 
In Figure~\ref{fig:atlaswele_njet}
we show the ratio of the predicted
theory cross-sections to the data for the ATLAS $W$+jets measurements
\cite{Aad:2010pg}. In this measurement jets are defined by the anti-Kt
algorithm \cite{Cacciari:2008gp}, with a radius parameter $R = 0.4$
and are required to have $\pt > 20~\rm{GeV}$ and $|\eta| <
2.8$.

The \alpgenpythiasix \pelfmatched prediction compares well with the measured
cross sections both as a function of the inclusive jet multiplicity and jet
$\pt$. In particular, the prediction correctly describes  the low $\pt$
region of the differential cross section and the jet sub-structure without
presenting any significant disagreement with data at high $\pt$. 
This shows that it is possible to tune separately the long- and short-distance 
contributions of the prediction to obtain a satisfactory description
of the observables in the whole experimental accessible phase space.   
Remarkably, the prediction describes the data both at $\sqrt{s}=1.96$~TeV and
 $\sqrt{s}=7$~TeV. This illustrates that the scaling properties of the long-distance
contribution with the centre of mass energy of the collision does not
produce unexpected effects in the high $\pt$ region of the cross section.
A coherent rescaling of \alphaS with \ktfac=0.5,~2 has a little effect on the
shapes of the differential cross-sections, while for the inclusive $N_{jet}$
cross-sections it produces variations that
bracket the default prediction. The \ktfac parameter can therefore be used to explore the
sensitivity of the prediction to a variation of the renormalization and
factorization scale, other than allow tuning on data.
With the statistics of the currently available measurements there is no much
room for optimising the parameters of the \alpgenpythiasix \pelfmatched
prediction in a tune that better describes the measurements. However,
with the imminent arrival of more precise higher-statistics LHC measurements, 
this should be possible in the near future.

Even though not explicitly shown here, in the relevant publications
one can find similarly good agreement between the available
measurements~\cite{Aaltonen:2007ip,ATLAS-CONF-2011-060,ATLAS-CONF-2011-042}
and predictions based on AlpGen + Herwig.

\subsection{Jets shapes}

Finally we test the ability of the new \alpgenpythiasix \pelfmatched tune and the
systematics variations to describe the jet shapes at the LHC and the Tevatron. 

For the LHC, the jet shapes measured in inclusive jet production 
by the ATLAS collaboration \cite{Aad:2011kq} are taken as
reference. For this measurement the jets are reconstructed using the
anti-kt algorithm with the distance parameter $R$=0.6, the transverse
momentum range 30~GeV $<\pt<$ 600~GeV and rapidity in the region
$\lvert y \rvert <$ 2.8. The jet shapes are expected to be sensitive
to both perturbative (parton shower) and non-perturbative
(fragmentation and underlying event) modelling aspects. We perform the data
comparisons for both the \alpgenpythiasix and \pythiasixstand
cases. The samples are generated using different \pythiasixstand and
\alpgenpythiasix parameter settings as follows: for \pythiasixstand
the \pelf and the associated systematics tunes \pelfradhi and
\pelfradlo are compared. The same tunes are also used for the
generating the \alpgenpythiasix distributions whereby the \LambdaQCD
values are always set to the same values in \alpgen (using \ktfac) and
\pythiasix (i.e. the \pelfmatched central settings and the systematic variations
around the central settings). The setups are compared to the integral jet shape distributions as measured in the data. The integral jet shape is defined as the average fraction of the jet $\pt$ that lies inside a cone of radius $r$ concentric with the jet cone \cite{Aad:2011kq}:
\begin{equation}
\Psi(r) = \frac{1}{\rm{N_{jet}}}\displaystyle\sum\limits_{i=1}^{\rm{N_{jet}}} \frac{\pt(0,r)}{\pt(0,R)},~0\leq r \leq R~.
\label{eq:Int_JShape}
\end{equation}
The sum is performed over all the $\rm{N_{jet}}$ jets in the kinematic region of interest.

\begin{figure}
\begin{center}
\begin{minipage}{150mm}
\includegraphics[width=0.5\textwidth]{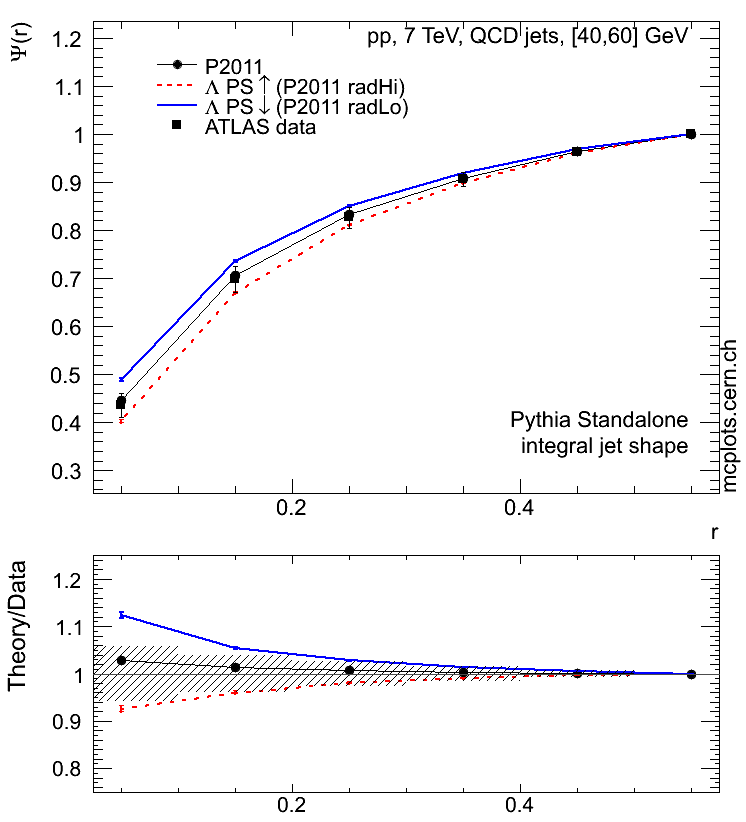}
\includegraphics[width=0.5\textwidth]{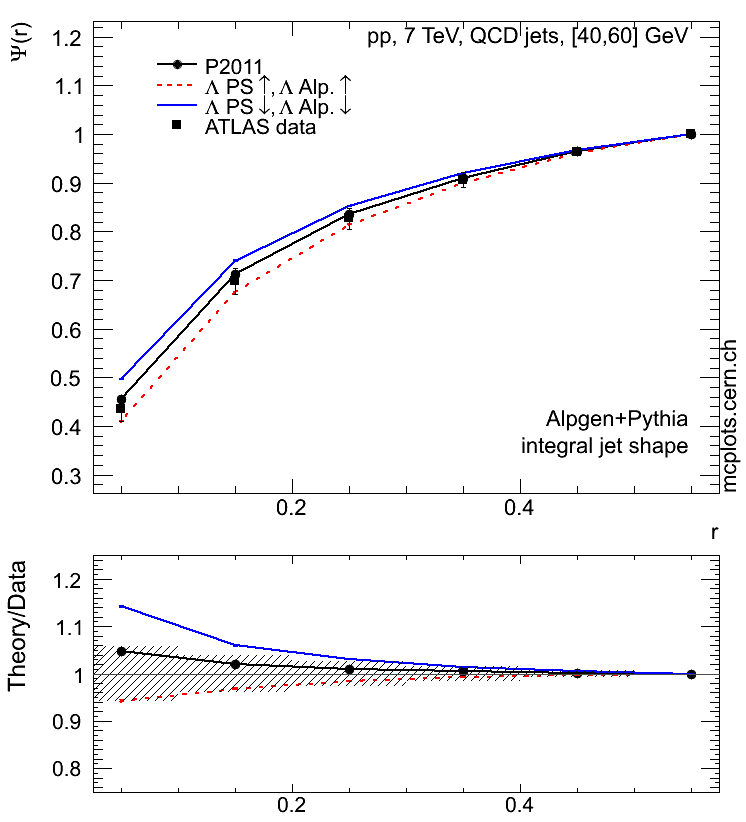}
\includegraphics[width=0.5\textwidth]{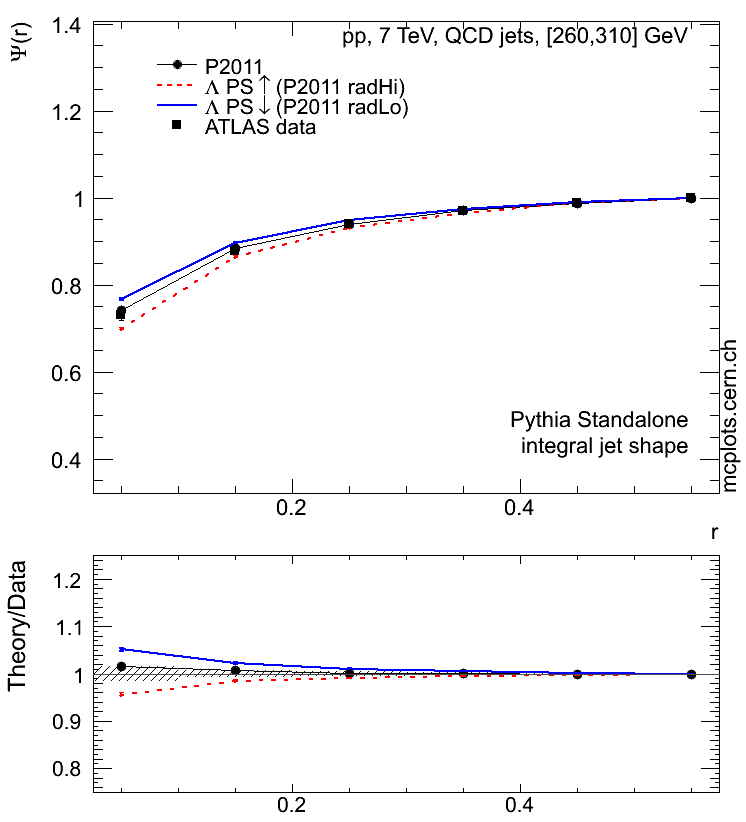}
\includegraphics[width=0.5\textwidth]{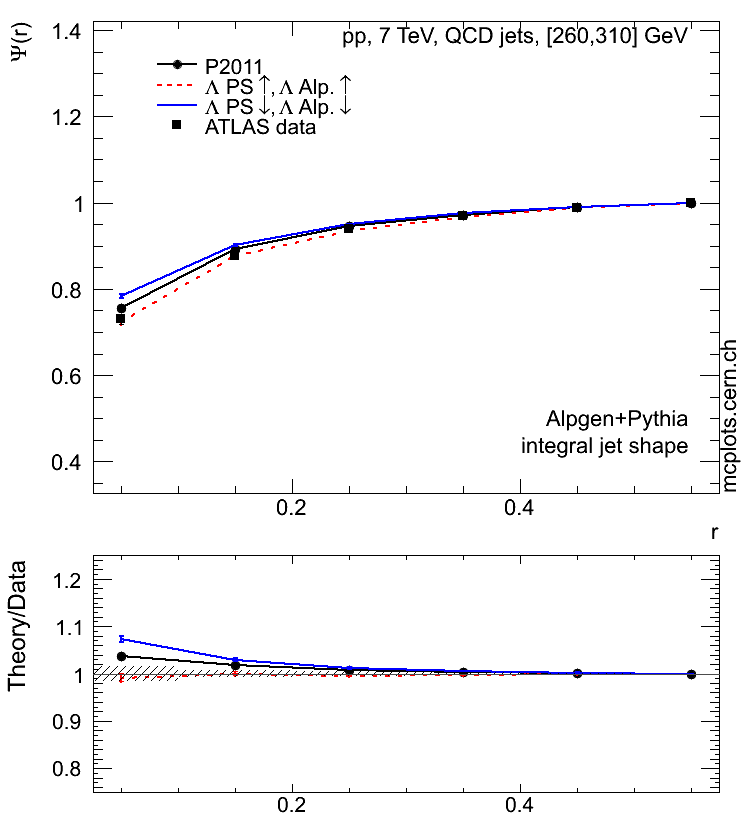}
\caption{Comparison of the integral jet shapes as measured by ATLAS
  \cite{Aad:2011kq} with the predictions of the \pythiasixstand (left)
  and \alpgenpythiasix (right) using \pelf, \pelfradhi and \pelfradlo
  tunes. The comparisons are performed for the jets with $|y|<2.8$ and
  $\pt$ ranges of 40-60~GeV (top) and 260-310~GeV (bottom).}
\label{fig:ATLAS_alpgen_jshape_30}
\end{minipage}
\end{center}
\end{figure}

In Figure \ref{fig:ATLAS_alpgen_jshape_30} the integral jet shape distributions are compared to the ATLAS 
data for the jets in the transverse momentum ranges of 40-60~GeV (top) and 260-310~GeV (bottom) in the whole measured rapidity range ($\lvert y \rvert <$ 2.8). 
We observe that both \pythiasixstand (left) and \alpgenpythiasix (right) with \pelf provide 
reasonably good description of the jet shapes. Due to MLM matching the jets in the \alpgenpythiasix 
case tend to be more narrow than in the \pythiasixstand case. 

\begin{figure}
\begin{center}
\begin{minipage}{150mm}
\includegraphics[width=0.85\textwidth]{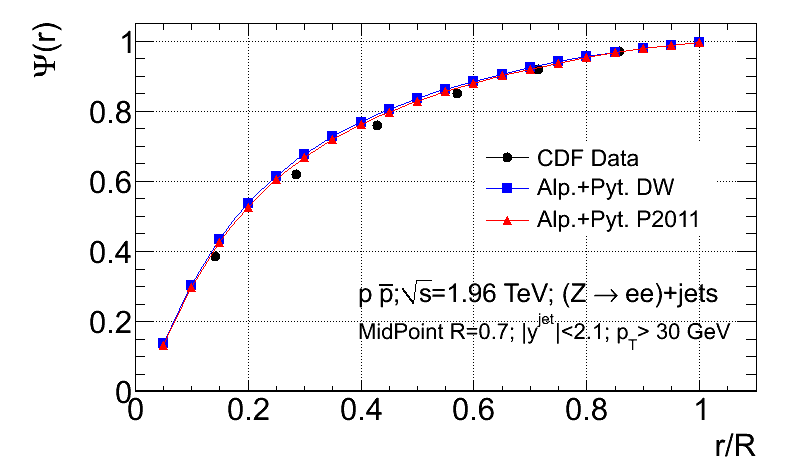}
\caption{A comparison of the predicted and CDF measured jet shape in
  $Z$+jets events, showing the fraction of the total energy of a jet
  of radius $R$ contained within a radius $r$, $\Psi(r)$, as a function of
  $r/R$. ``Alp.+Pyt. DW'' uses the default \alpgen parameters and the DW \pythia tune, ``Alp.+Pyt. P2011'' is described in Section ~\ref{sec:newtune} and it includes a consistent choice of \lamqcd~ in \alpgen and \pythia. }
\label{fig:cdfz_jetshape}
\end{minipage}
\end{center}
\end{figure}

For the Tevatron, the shapes of jets produced 
in association with a Z boson as measured by CDF \cite{note:cdfJS} are used. 
In this measurement jets are defined by the CDF midpoint
algorithm \cite{Abulencia:2005yg}, with $R_{cone} = 0.7$
and are required to have $p^{jet}_{T} > 30~\rm{GeV}$ and $|y^{jet}| <
2.1$. Figure~\ref{fig:cdfz_jetshape}  shows good agreement between
\alpgenpythiasix and the measurement for both the DW and \pelf tunes.

The comparisons in Figures~\ref{fig:ATLAS_alpgen_jshape_30} and \ref{fig:cdfz_jetshape} 
(as well as comparisons to the jet shapes in other kinematic regions and comparisons to further LHC measurements) reveal no
major short-comings of the \alpgenpythiasix \pelfmatched tune. 
The \pelf tune has been developed by tuning \pythiasixstand, whereby the effective value of \lamqcd~ was set to be the same throughout
the \pythiasix~parton shower in the anticipation of using it with the \alpgen matrix elements using the same 
effective \lamqcd~ value. The agreement with the measured jet shapes data could therefore
potentially be improved by performing a dedicated tuning of \alpgenpythiasix.

\section{Conclusions\label{sec:conclusion}}
We have shown that, in the context of tuning ME-PS matched
predictions, it is vital that the tuning adopted ensures a consistent
treatment of \alphaS~on either side of the ``matching boundary''. In
the case of \alpgenpythiasix matched predictions, we have outlined a
simple prescription to ensure this. This can be easily generalised, and
applied to the case of matching \alpgen with other shower MCs, such as
\herwig. We have then given an example of such a
tune that compares well to Tevatron and LHC measurements of vector
boson plus multijet final states. In addition, we have shown how
consistent variations around a central ME-PS matched tune can be
performed, so as to define a systematic uncertainty on that
prediction. This knowledge should prove valuable in defining a new set
of consistent ME-PS tunes for the precise future study of LHC multijet
final states.

\subsection*{Acknowledgements}
J.~K. and L.~M. acknowledge the support of the Initiative and
Networking Fund of the Helmholtz Association, contract HA-101
("Physics at the Terascale"). B.~C. acknowledges the support of the UK
Science and Technology Facilities Council (STFC). L.~M. acknowledges
the LPCC for the hospitality during the initial phase of this project.

\clearpage
\appendix
\section{Appendix}

\subsection{An \alphaS~Consistent
  \alpgenpythiasix~Tune\label{sec:aptune}}

We here describe the parameters in the
\alpgen~and \pythia 6 codes that are important for ensuring the
consistency of matching, and give the settings of these parameters
which describe a new \alphaS~consistent \alpgenpythiasix~tune.
Note that the settings for \pythia~are those of the central
Perugia-2011 tune~\cite{Skands:2010ak}, which was inspired by these
studies. 

\subsubsection{\alpgen Parameters}
For \alpgen, the relevant parameters controlling the magnitude,
renormalization scale, and running order for \alphaS
are set to these values:\footnote{The parameters \xlclu and \ttt{lpclu} were
  introduced in connection with this work, and are implemented in
  \alpgen starting from v2.14. The default behaviour, if these
  parameters are not set, is to assign the values inherited from the PDF (as
  was the case for \alpgen versions before 2.14). }:
\begin{itemize}
\item The prefactor rescaling the value of the renormalization scale
  in the CKKW scale-setting procedure: \ktfac = 1.0
\item \LambdaQCD (5-flavour): \xlclu =0.26
\item Running order: \ttt{lpclu} = 1
\end{itemize}

\subsubsection{\pythia Parameters\label{sec:pythparams}}
To the extent that the explanations given here are
necessarily somewhat brief, we recommend the
interested reader to follow up on the definitions of the parameters
below in the program's comprehensive manual
\cite{Sjostrand:2006za}. 

\paragraph{The Strong Coupling in Pythia 6:} there are several different ways of specifying the
parameters controlling \alphaS. 
In order for the user to have explicit control of all of them, 
we use the option \ttt{MSTP(3)=1}, which allows to
specify formally independent \LambdaQCD values for each of the
different algorithmic components in \pythia. As a special case, we may
then choose to set all those values equal, or at least set those equal that
correspond to initial- and final-state radiation. These
are:
\begin{itemize}
\item \ttt{PARP(61)} for \LambdaQCD for ISR.
\item \ttt{PARJ(81)} for \LambdaQCD for FSR inside resonance decays. 
\item \ttt{PARP(72)} for \LambdaQCD for FSR outside resonance decays (e.g.,
  FSR off hard jets from the matrix element and/or from ISR).
\end{itemize}
The number of flavours with which to
interpret \LambdaQCD must also be specified. 
Since a value of $n_f=5$ is hard-coded
in at least one sub-algorithm of \pythia, we advise to always translate
\LambdaQCD values to 5-flavour ones and, correspondingly, set the
$n_f$ parameter \ttt{MSTU(112)=5}. 

Further, in order to avoid that any of these values are modified by
the code, we set \ttt{MSTP(64)=2}, and \ttt{PARP(64)=1.0}. The former 
forces the code to keep \LambdaQCD\ unmodified for ISR. 
In particular, the translation from ``MSbar'' to ``CMW'', which is
applied for \ttt{MSTP(64)=3}, is \emph{not} performed. This is 
equivalent to interpreting the effective \LambdaQCD\ value as already
being in a scheme similar to CMW. The latter, \ttt{PARP(64)=1.0}, 
sets the prefactor for the renormalization scale used for ISR 
equal to unity, i.e., the renormalization scale will just be
$\pt$. 
Any re-interpretation of \LambdaQCD, for instance to translate
between different effective scheme definitions 
or to introduce multiplicative factors on the effective renormalization scale, for
    scale variation purposes, 
    should then be imposed directly on the three \LambdaQCD 
    values above. This is the prescription followed in the so-called
    Perugia 2011 tunes which were developed as part of this effort,
    with parameters as listed in \cite{Skands:2010ak}. 

Finally, one needs to settle on an effective \emph{value} for
\LambdaQCD. According to  comprehensive \professor tunings  
\cite{Buckley:2009bj, AtlasTuneNote} of the $\pt$-ordered shower in \pythia
\cite{Sjostrand:2004ef} to event shapes and other LEP data, 
one needs values of order 
\begin{equation}
\Lambda^{(5)}_{\mrm{QCD}} \sim 0.26~,
\end{equation}
where the superscript indicates the number of flavours. We interpret
this as an effective value, derived directly from data using a
 ``\pythia scheme'' that is defined numerically by \pythia's shower
algorithm. It is not necessarily directly comparable to 
$\overline{\mrm{MS}}$ determinations\footnote{It is probably closest to the
so-called CMW scheme \cite{Catani:1990rr}. 
For completeness, a reasonable derived guess for a
corresponding $\overline{\mrm{MS}}$ value would then be $\sim 0.16$,
since the CMW prescription yields effective values that are 
approximately 1.6 times larger than the $\overline{\mrm{MS}}$ one.}.

\paragraph{The Strong Coupling in Pythia 8:} 
Although we restrict the numerical studies in this paper 
to \pythiasix, for
completeness we also include
the case of \pythia~8 \cite{Sjostrand:2007gs}, for which 
the corresponding relevant parameters are
\begin{itemize}
\item  \ttt{TimeShower:alphaSvalue}, 
\item  \ttt{TimeShower:alphaSorder},
\item  \ttt{SpaceShower:alphaSvalue}, 
\item \ttt{SpaceShower:alphaSorder},
\end{itemize}
for final-state (timelike) and initial-state (spacelike) showers,
respectively. Notice in particular that one here 
specifies the value of
$\alpha_s(M_Z)$ rather than that of \LambdaQCD. Similar
comments about the effective scheme definition as 
for \pythiasix apply. 

\paragraph{Radiation Phase Space:} 
The size of the allowed phase space for radiation in the shower
generator may also affect the matched result. 
In the $\pt$-ordered shower in \pythiasix, the switch
\ttt{MSTP(72)} controls the starting scale for final-state radiation
off jets that are produced by initial-state radiation and/or are
colour-connected to the beam. Naively, the FSR off such a parton
should start at the scale at which it was created, which is obtained
with \ttt{MSTP(72)=2}, the recommended option. Using the other available
options is strongly discouraged, as these lead to a quite bad
agreement with NLL resummations, underscored by Banfi et al in
\cite{Banfi:2010xy}. The Perugia 2011 tunes all use \ttt{MSTP(72)=2}. 

Further, in the Perugia
2011 tunes, we start both the ISR and FSR evolutions at $p_{T\mrm{evol}} =
\mbox{\ttt{SCALUP}}$, with the \pythia evolution variable
$p_{T\mrm{evol}}$ defined in \cite{Sjostrand:2004ef} and
\ttt{SCALUP} the scale parameter defined in the Les Houches Accord for
event generators \cite{Boos:2001cv,Alwall:2006yp}. Technically, this
is achieved by setting \ttt{PARP(71) = PARP(67) = 1.0}, where the
former controls the scale factor applied to the starting scale for FSR
and the latter sets the one for ISR.
Note that these
parameters could still be varied somewhat around their central values,
since the  $p_{T\mrm{evol}}$ variable used by \pythia is not
100\% identical to the $\pt$ definition that might be used to
place cuts in a matrix-element generator, but we have not judged this
difference essential at the current level of precision.

\subsection{PYTHIA Tunes \label{sec:tunes}}
In this work, we have used the Perugia 0, Perugia Hard, Perugia Soft,
Perugia 2010, Perugia 2011, Perugia 2011 radHi, Perugia 2011
radLo~\cite{Skands:2010ak}, and the DW~\cite{Albrow:2006rt} tunes of 
\pythia~6.4~\cite{Sjostrand:2006za}. All use the CTEQ5L PDF set~\cite{Lai:1999wy}.
For a complete description, see
the indicated references. The salient features of the tunes are as
follows. 

Tune DW~\cite{Albrow:2006rt} is a tune of the $Q^2$-ordered shower. It
is based on the 
Tevatron ``Tune A'', which had great success in describing the
underlying event  measured at the Tevatron. Contrary to Tune A, 
however, DW also included the Drell-Yan $\pt$ spectrum, for which
Tune A predicted a far too soft spectrum. Tune DW therefore has a
significantly lower renormalization scale for ISR (and thus a larger
value of $\alpha_s$), and 2~GeV of so-called ``primordial
$k_\perp$'', as compared to 1~GeV in Tune A. The energy scaling of the
underlying event was based on comparisons between the underlying-event
level at the Tevatron between 630 and 1800~GeV.

The Perugia tunes~\cite{Skands:2010ak} are all tunes of the $\pt$-ordered
shower. Unlike DW, which was developed by tuning to the underlying
event in jet events, the Perugia tunes primarily used minimum-bias
data as drivers, relying on the universality of PYTHIA's MPI modelling
to extrapolate to the underlying event. In addition, a comprehensive
update of the LEP fragmentation parameters was included in all tunes. 

The first set, the Perugia 0 family, used LEP event shapes and
fragmentation data, Tevatron minimum-bias data, and the Tevatron Drell-Yan $\pt$
spectrum. Again, the scaling from Tevatron data at 
630~GeV was used to determine the scaling with CM energy, with some
additional constraints from older UA5 data also included. A ``Hard''
and ``Soft'' variation attempted to vary the 
shower radiation up and down, respectively. Both Perugia 0 and the
``Hard'' variation use the so-called ``CMW'' scheme for
\LambdaQCD\ for ISR, while the soft retained the unmodified MSbar value, in all
cases taking the numerical value from the PDF set used. For the
``Hard'' variation, the renormalization scale for ISR was
$0.5\pt$, for Perugia 0 $\pt$, and for the ``Soft'' variation,
$\sqrt{2}\pt$. In addition, the ``Hard'' variation had 
higher-than-nominal values for FSR, and had a slightly harder
hadronization spectrum, while the converse was true for the ``Soft''
one. None of these early tunes used the recommended \ttt{MSTP(72)=2}
setting, and hence predicted rather narrow ISR jets. The \pythia\ tune
numbers are 320, 
321, and 322, for Perugia 0, ``Soft'', and ``Hard'', respectively.

Tab.\ref{PerugiaTable} lists  the  parameter settings of the Perugia family that were used for the block variations in \ref{sec:example} .
\begin{table}[t]
\begin{center}
\begin{tabular}{llccc}
\toprule
 tuning block & parameter or switch &  Perugia 0  & Perugia hard & Perugia soft\\ 
\midrule
ISR   & PARP(64)  & 1.0 & 0.25 & 2.0 \\
ISR  & PARP(67) & 1.0 & 4.0 & 0.25\\
ISR  & MSTP(64) & 3 & 3 & 2 \\ \hline
FISR & PARP(71) & 2.0 & 4.0 & 1.0 \\
FISR & MSTP(72) & 1 & 1 & 0 \\ \hline
UE & PARP(82) & 2.0 & 2.3 & 1.9 \\
UE & PARP(83) & 1.7 & 1.7 & 1.5 \\
UE & PARP(90) & 0.26 & 0.30 & 0.24 \\ \hline
CR & PARP(77) & 0.9 & 0.4 & 0.5 \\
CR & PARP(78) & 0.33 & 0.37 & 0.15 \\
\bottomrule
\end{tabular}
\caption{ Table of Perugia tune parameters  relevant for this study. For a complete list see  \cite{Skands:2010ak} }
\label{PerugiaTable}
\end{center}
\end{table}

In Perugia 2010, jet shapes were included among the tuning
constraints. The amount of FSR outside resonance decays (previously
controlled by the \LambdaQCD\ value read from the PDF set) was
adjusted to agree with the level inside them (constrained by fits to
LEP event shapes), combined with the recommended \ttt{MSTP(72)=2}.
The \LambdaQCD\ value for ISR was still read from
the PDF set, and translating from MSbar to CMW, as in Perugia 0. 
A few fragmentation parameters were slightly
revised, since some of the previous ones had only been constrained using the
$Q^2$-ordered shower, and a new colour-reconnection model was
introduced. No ``Hard'' and ``Soft'' variations were produced for this
tune. The  \pythia\ tune number is 327 for Perugia 2010.

In Perugia 2011, it was possible to include some early
lessons from LHC at 7 TeV. Based on observed strangeness and baryon
production rates, a few of the fragmentation parameters were again
revised. The universal effective \LambdaQCD\ choice advocated in this
paper was introduced. Variations labelled ``radHi'' and ``radLo'' were
defined as well, expressing a factor 2 variation in the \LambdaQCD\ values
used for ISR and FSR. The \pythia\ tune numbers for Perugia 2011,
radHi, and radLo, are 350, 351, and 352, respectively.

Tabulated values of the parameters of all of the Perugia tunes can be found
in the appendices of \cite{Skands:2010ak}.

\clearpage
\bibliographystyle{h-physrev5}
\bibliography{main}

\end{document}